\documentclass[10pt,preprint]{emulateapj}  

\usepackage{epsfig}
\usepackage{subfigure,verbatim}

\newcommand{\unit}[1]{\nobreak{\mathrm{\;#1}}} 

\newcommand{\cross}{\times}
\newcommand{\bmath}[1]{\mbox{\boldmath{$#1$}}}

\newcommand{\bi}{\begin{itemize}}
\newcommand{\ei}{\end{itemize}}
\newcommand{\tit}[1]{\textit{#1}}

\newcommand{\ex}[1]{10^{-#1}}
\def\comp{\,c/\omega_{\rm pi}}
\def\compe{\,c/\omega_{\rm pe}}
\def\ompt{\omega_{\rm pi}t}

\def\mime{m_i/m_e}

\newcommand{\eq}[1]{eq.~(\ref{eq:#1})}
\newcommand{\fig}[1]{Fig.~\ref{fig:#1}}
\newcommand{\be}{\begin{eqnarray}}
\newcommand{\ee}{\end{eqnarray}}

\begin{document}
\title{The Maximum  Energy of Accelerated Particles in Relativistic Collisionless Shocks}
\author{Lorenzo Sironi,$^{1,2}$ Anatoly Spitkovsky,$^3$ and Jonathan Arons$^4$}
\affil{$^1$ Harvard-Smithsonian Center for Astrophysics, 
60 Garden St., Cambridge, MA 02138, USA; lsironi@cfa.harvard.edu
\\
$^2$ NASA Einstein Post-Doctoral Fellow
\\
$^3$ Department of Astrophysical Sciences, Princeton University, Princeton, NJ 08544-1001, USA
\\
$^4$ Department of Astronomy, Department of Physics, and Theoretical Astrophysics Center, University of California, Berkeley, CA 94720}
 
\begin{abstract}
The afterglow emission from gamma-ray bursts (GRBs) is usually interpreted as synchrotron radiation from electrons accelerated at the GRB external shock, that propagates with relativistic velocities into the magnetized interstellar medium. By means of multi-dimensional particle-in-cell simulations, we investigate the acceleration performance of weakly magnetized relativistic shocks, in the magnetization range $0\lesssim\sigma\lesssim\ex{1}$. The pre-shock magnetic field is orthogonal to the flow, as generically expected for relativistic shocks. We find that relativistic perpendicular shocks propagating in electron-positron plasmas are efficient particle accelerators if the magnetization is $\sigma\lesssim\ex{3}$. For electron-ion plasmas, the transition to efficient acceleration occurs for $\sigma\lesssim 3\times 10^{-5}$. Here, the acceleration process proceeds similarly for the two species, since the electrons enter the shock nearly in equipartition with the ions, as a result of strong pre-heating in the self-generated upstream turbulence. In both electron-positron and electron-ion shocks, we find that the maximum energy of the accelerated particles scales in time as $\varepsilon_{max}\propto t^{1/2}$. This scaling is shallower than the so-called (and commonly assumed) Bohm limit $\varepsilon_{max}\propto t$, and it naturally results from the small-scale nature of the Weibel turbulence generated in the shock layer. In magnetized plasmas, the energy of the accelerated particles increases until it reaches a saturation value $\varepsilon_{sat}/\gamma_0 m_i c^2\sim\sigma^{-1/4}$, where $\gamma_0 m_i c^2$ is the mean energy per particle in the upstream bulk flow. Further energization is prevented by the fact that the self-generated turbulence is confined within a finite region of thickness $\propto \sigma^{-1/2}$ around the shock. Our results can provide physically-grounded inputs for models of non-thermal emission from a variety of astrophysical sources, with particular relevance to GRB afterglows.
\end{abstract}


\section{Introduction}\label{sec:intro}
The external shocks of gamma-ray bursts (GRBs) are often invoked as efficient sites of acceleration for protons and electrons \citep[e.g.,][]{waxman_06}. Shock-acceleration of protons might explain the flux of Ultra High Energy Cosmic Rays (UHECRs) observed with energies in excess of $10^{20}\unit{eV}$ by the Pierre Auger Observatory \citep[][]{auger_10}. Synchrotron emission from the shock-accelerated electrons powers the GRB afterglow emission, which is usually detected in the X--ray, optical and sometimes radio bands, and recently up to sub-GeV energies by the Fermi telescope \citep[e.g.,][]{ackermann_10,depasquale_10,ghisellini_10}.

The external shocks in GRB afterglows are believed to be relativistic shocks propagating in weakly magnetized electron-proton plasmas, either the interstellar medium (ISM) or the wind of the progenitor star. If the ISM number density is $n\equiv n_{0}\unit{cm^{-3}}$ and the magnetic field is $B_{\rm ISM}\equiv 3\,B_{{\rm ISM},-5.5}\unit{\mu G}$, the ISM magnetization will be
\be\label{eq:sigma}
\sigma=\frac{B_{\rm ISM}^2}{4\pi n m_i c^2}\simeq0.5\times\ex{9} B_{{\rm ISM},-5.5}^2 n_{0}^{-1}~,
\ee
where $m_i$ is the proton mass and $c$ is the speed of light. For relativistic shocks, the mean field in the post-shock frame will be mostly transverse to the flow, due to shock compression and to the effect of the Lorentz transformations. For perpendicular shocks (i.e., with the field orthogonal to the flow), the magnetization in \eq{sigma} is a Lorentz invariant, and independent of the shock radius (provided that $n$ and $B_{\rm ISM}$ are constant in radius).\footnote{The independence of $\sigma$ from the shock radius also holds for a wind profile of the external density, under the assumption that the field is primarily toroidal.}

The acceleration process at the external shocks of GRBs is thought to be governed by the Fermi mechanism, where particles stochastically diffuse back and forth across the shock front and gain energy by scattering from magnetic turbulence embedded in the converging flows \citep[e.g.,][]{blandford_ostriker_78, bell_78,drury_83,blandford_eichler_87}. The highly nonlinear physics of the Fermi process -- where the magnetic turbulence that mediates the particle acceleration is generated by the particles themselves -- can only be addressed self-consistently by means of first principle particle-in-cell (PIC) simulations. By using PIC simulations, \citet{sironi_spitkovsky_09} (in electron-positron flows; hereafter, SS09) and \citet{sironi_spitkovsky_11a} (in electron-ion flows; hereafter, SS11) have demonstrated that particle acceleration is suppressed in $\sigma\sim0.1$ perpendicular shocks. Here, the self-generated turbulence is not strong enough to permit efficient injection into the Fermi process. On the other hand, in the extreme case of unmagnetized shocks (i.e., $\sigma=0$), PIC simulations have shown that a non-thermal tail of shock-accelerated particles is generated, as a self-consistent by-product of the shock evolution	\citep{spitkovsky_08,spitkovsky_08b,martins_09,haugbolle_10}.

In this work, we investigate the physics of weakly magnetized perpendicular shocks for magnetizations in the range $0\lesssim \sigma\lesssim\ex{1}$. We explore how the acceleration efficiency, namely the fraction of particles and energy stored in the non-thermal tail, depends on the flow magnetization, in electron-positron (\S\ref{sec:pairs}) and electron-ion (\S\ref{sec:ions}) flows. We show that shocks propagating in electron-positron plasmas are efficient particle accelerators, if the magnetization is $\sigma\lesssim\ex{3}$. A smaller threshold is found in the case of electron-ion plasmas ($\sigma\lesssim 3\times 10^{-5}$). Here, the acceleration process proceeds similarly for the two species, since the electrons enter the shock nearly in equipartition with the ions, as a result of strong pre-shock heating in the self-generated turbulence.

In all the cases where the Fermi process is efficient, we follow the temporal evolution of the upper cutoff of the non-thermal tail of accelerated particles, finding that the maximum energy scales as $\varepsilon_{max}\propto t^{1/2}$, for both electron-positron and electron-ion flows. This is in contrast with the \textit {ad hoc} prescription given by the so-called Bohm scaling (namely, $\varepsilon_{max}\propto t$), which is often employed in the literature, for lack of a better choice. In magnetized plasmas, we find that the energy of the accelerated particles increases up to $\varepsilon_{sat}/\gamma_0 m_i c^2\sim\sigma^{-1/4}$, where $\gamma_0 m_i c^2$ is the mean energy per particle in the upstream bulk flow. Further energization is prevented by the fact that the self-generated turbulence is confined within a finite region of thickness $\propto \sigma^{-1/2}$ around the shock. 

With our PIC simulations, we are then able to provide a physically-grounded scaling for the acceleration rate in weakly magnetized relativistic perpendicular shocks. Our results can be easily incorporated in models of non-thermal emission from astrophysical sources, as we illustrate in \S\ref{sec:apply}. We finally summarize our findings in \S\ref{sec:disc}.



\begin{figure}[tbp]
\begin{center}\label{fig:simplane}
\includegraphics[width=0.5\textwidth]{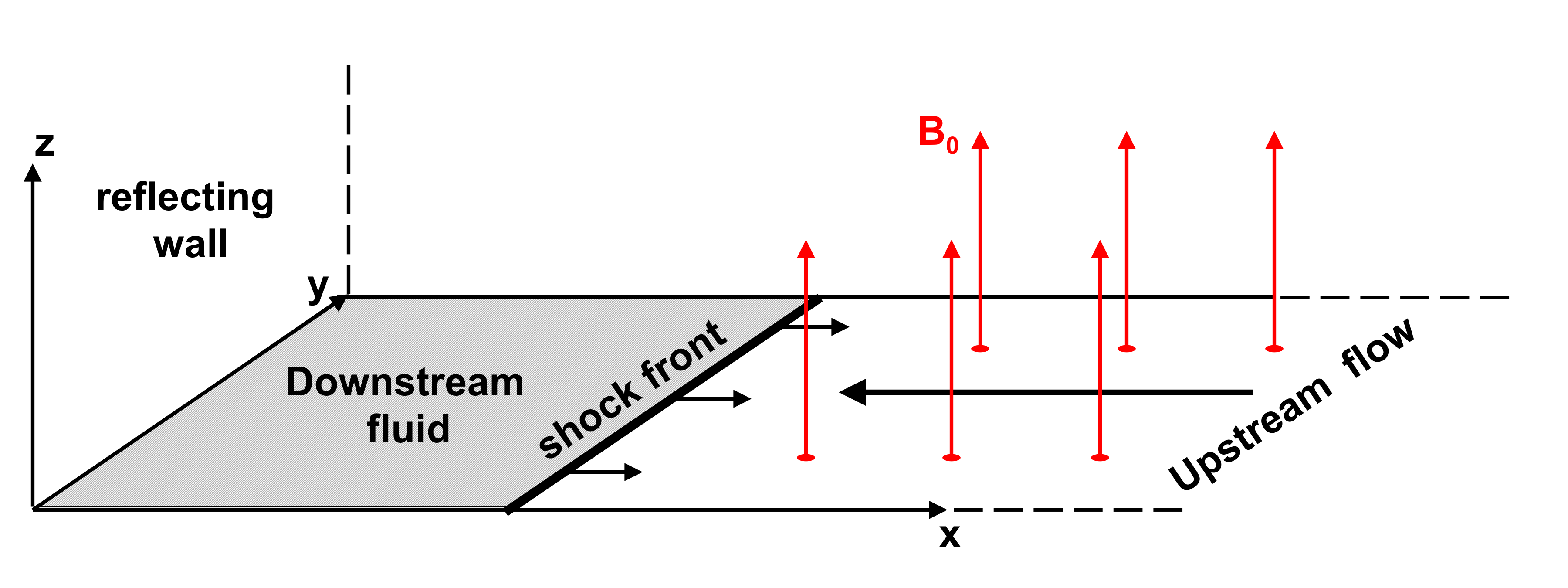}
\caption{Simulation geometry. For our 2D simulations,  the computational domain is in the $xy$ plane, with periodic boundary conditions in the $y$ direction. The incoming flow propagates along $-\bmath{\hat{x}}$, and the shock moves away from the reflecting wall (located at $x=0$) toward $+\bmath{\hat{x}}$. The magnetic field carried by the upstream flow (red arrows) is perpendicular to the simulation plane.}
\end{center}
\end{figure}

\section{Simulation Setup}\label{sec:setup}
We use the three-dimensional (3D) electromagnetic PIC code TRISTAN-MP \citep{spitkovsky_05}, which is a parallel version of the public  code TRISTAN \citep{buneman_93} that was optimized for studying relativistic collisionless shocks. Our simulation setup parallels SS09 and SS11 very closely, which we repeat here for completeness.

The shock is set up by reflecting a cold ``upstream''  flow from a conducting wall located at $x = 0$ (\fig{simplane}). The interaction between the incoming beam (that propagates along $-\bmath{\hat{x}}$) and the reflected beam triggers the formation of a shock, which moves away from the wall along $+\bmath{\hat{x}}$. This setup is equivalent to the head-on collision of two identical plasma shells, which would form a forward and reverse shock and a contact discontinuity. Here, we follow only one of these shocks, and replace the contact discontinuity with the conducting wall. 
The simulation is performed in the ``wall'' frame, where the ``downstream'' plasma behind the shock is at rest. 

We perform simulations in both 2D and 3D computational domains, and we find that most of the shock physics is well captured by 2D simulations. Therefore, to follow the shock evolution for longer times with fixed computational resources, we mainly utilize 2D runs, but we explicitly show that our 2D results are in excellent agreement with large 3D simulations. For both 2D and 3D domains, all three components of particle velocities and electromagnetic fields are tracked. In our 2D simulations, we use a rectangular simulation box in the $xy$ plane, with periodic boundary conditions in the $y$ direction (\fig{simplane}). In 3D, we employ periodic boundary conditions both in $y$ and in $z$. Each computational cell is initialized with four particles (two per species) in 2D and with two particles  (one per species) in 3D. We have performed limited experiments with a larger number of particles per cell (up to 8 per species in 2D), obtaining essentially the same results.

We investigate the physics of both electron-positron and electron-ion shocks. In both cases, the relativistic  skin depth for the incoming electrons ($c/\omega_{\rm pe}$)  is resolved with 8 computational cells, and the simulation timestep is $\Delta t=0.056\,\omega_{\rm pe}^{-1}$. Here, $\omega_{\rm pe}\equiv(4\pi e^2 n_{e} /\gamma_0 m_e)^{1/2}$ is the relativistic  plasma frequency for the upstream electrons, with number density $n_{e}$ (measured in the wall frame) and bulk Lorentz factor $\gamma_0$. For electron-positron flows, the  plasma frequency of the upstream positrons  is $\omega_{\rm pi}=\omega_{\rm pe}$. In the case of electron-ion shocks, we typically  employ a reduced mass ratio $m_i/m_e=25$, which allows to follow the shock evolution for sufficiently long times (in units of the inverse ion plasma frequency $\omega_{\rm pi}^{-1}=\sqrt{m_i/m_e}\,\omega_{\rm pe}^{-1}$), while still clearly separating the ion and electron dynamical scales.\footnote{We remark that in the following we use the same symbol $\omega_{\rm pi} $ to indicate the relativistic plasma frequency of positrons (in electron-positron flows) and of ions (in electron-ion shocks). Similarly, we indicate with $m_i$ the mass of the positively-charged particles (positrons or ions, depending on the flow composition).} As we show in Appendix \ref{sec:specmime}, we obtain essentially the same results when using higher mass ratios (we have tried up to $m_i/m_e=1600$,  approaching the realistic value $m_i/m_e\simeq1836$), which suggests that a mass ratio $m_i/m_e=25$ is already ``large'' enough to capture the acceleration physics in our electron-ion shocks. 

For electron-positron shocks, our computational domain in 2D is typically $\sim128\comp$ wide (corresponding to 1024 cells), and in 3D the transverse size of the box amounts to $\sim64\comp$, or 512 cells. We have tried with 2D boxes up to three times as wide, finding essentially the same results. In electron-ion shocks, we  choose a box with a width of 1024 cells in 2D and 512 cells in 3D, which correspond respectively to $\sim26\comp$ and $\sim13\comp$, for our reference mass ratio $m_i/m_e=25$. When scaling to higher mass ratios, we choose a 2D computational domain with 1024 transverse cells for $\mime=100$, with 2048 cells for $\mime=400$, and with 4096 cells for $\mime=1600$, so that in each case the width of the box amounts to $\sim13\comp$. 

The incoming plasma is injected through a ``moving injector,'' which recedes from the wall along $+\bmath{\hat{x}}$ at the speed of light. The simulation box is expanded in the $x$ direction as the injector approaches the right boundary of the computational domain. This permits us to save memory and computing time, while following the evolution of all the upstream regions that are causally connected with the shock. The final time of our simulations is chosen such that we can confidently predict the subsequent evolution of the shock. Our longest 3D simulation ran for 50,000 timesteps, corresponding to a box with $22,500$ cells along $x$, while in 2D we were able to evolve a computational domain with $3072$ transverse cells up to $400,000$ timesteps, at which point the box extends over $180,000$ cells in the $x$ direction.

The cold incoming stream is injected along $-\bmath{\hat{x}}$ with bulk Lorentz factor $\gamma_0=15$. As we show in \S\ref{sec:pairs} and \S\ref{sec:ions}, our results are nearly the same for higher values of the Lorentz factor (we have explored from $\gamma_0=3$ up to $\gamma_0=240$), modulo an overall shift in the energy scale. The upstream flow is seeded with a background magnetic field $B_0$, which we parameterize in terms of the ratio of magnetic to kinetic energy density $\sigma\equiv B_{0}^2/4 \pi \gamma_0 m_i n_{i} c^2$, where $n_i$ (=$n_e$) is the number density of incoming positrons or ions.\footnote{In electron-ion shocks, the magnetization parameter as defined above pertains to ions. For electrons, $\sigma_{e,0}=(m_i/m_e)\,\sigma$ at injection, where the electron kinetic energy is $\gamma_0m_ec^2$. So, it seems that the electron magnetization would depend on $m_i/m_e$, for fixed $\sigma$. However, on their way to the shock, electrons increase their average energy up to a fraction $\epsilon_e$ of the initial ion energy $\gamma_0 m_i c^2$, where $\epsilon_e$ is nearly insensitive to $m_i/m_e$. So, the ``effective'' electron  magnetization $\sigma_{e,\rm{eff}}=(1/\epsilon_e)\,\sigma$ is insensitive to the mass ratio.} We investigate the regime of weakly magnetized flows, with magnetizations lower than the $\sigma=\ex{1}$ case explored by SS09 and SS11, down to the limit of unmagnetized shocks discussed by \citet{spitkovsky_08,spitkovsky_08b}. The regime of magnetizations we study here may  be relevant for external shocks in GRB afterglows, as we discuss in \S\ref{sec:grb}.

The angle between the shock direction of propagation $+\bmath{\hat{x}}$ and the upstream magnetic field $\mathbf{B}_0$ is taken to be $\theta=90^\circ$ in all cases, i.e., we only focus on ``perpendicular'' shocks (\fig{simplane}). As discussed in \S\ref{sec:intro}, this is the most relevant configuration for the relativistic shocks of GRB afterglows  in the frame of the post-shock medium, due to Lorentz transformation effects and shock compression. We refer to SS09 and SS11 for a complete investigation of the dependence of the shock physics on the field obliquity $\theta$, from $\theta=0^\circ$, which corresponds to a parallel shock, with magnetic field aligned with the shock normal, up to $\theta=90^\circ$, i.e., a perpendicular shock, with magnetic field along the shock front. In the upstream medium, we also initialize a motional electric field $\mathbf{E}_0=-\mbox{\boldmath{$\beta$}}_0\cross\mathbf{B}_0$, where  $\mbox{\boldmath{$\beta$}}_0=-\beta_0\,\bmath{\hat{x}}$ is the three-velocity of the injected plasma. In our 2D experiments, we choose a magnetic field orthogonal to the simulation plane, as shown in \fig{simplane}. By comparison with 3D simulations, we have verified that this is the field geometry most appropriate to capture the 3D physics of the shock, in the regime of magnetizations and Lorentz factors we investigate. In Appendix \ref{sec:specphi}, we show how our results depend on the orientation of the field with respect to the simulation plane.


\section{Electron-Positron Shocks}\label{sec:pairs}
In this section, we explore the physics of electron-positron shocks ($\mime=1$), focusing on the efficiency and rate of particle acceleration. In \S\ref{sec:pairnosig}, we discuss the long-term evolution of particle acceleration in unmagnetized shocks ($\sigma=0$), and in \S\ref{sec:pairsig} we  explore the physics of weakly magnetized flows, with $10^{-5}\lesssim\sigma\lesssim10^{-1}$. Finally, in \S\ref{sec:gamma} we investigate the dependence of our results on the pre-shock bulk Lorentz factor $\gamma_0$, for both unmagnetized and weakly magnetized shocks.

\begin{figure*}[tbp]
\begin{center}
\includegraphics[width=1\textwidth]{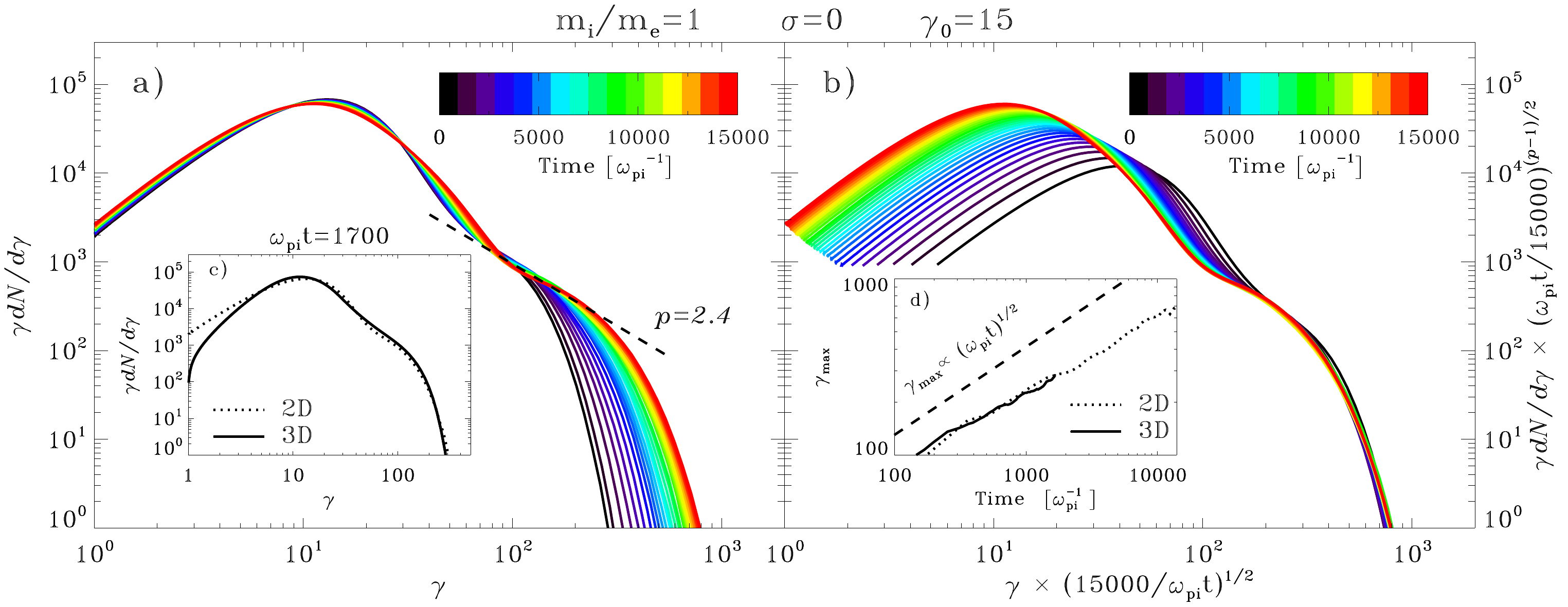}
\caption{Left panel: Temporal evolution of the post-shock particle spectrum, from the 2D simulation of a $\gamma_0=15$ electron-positron unmagnetized shock. We follow the evolution of the shock from its birth (black curve) up to $\ompt=15000$ (red curve). The non-thermal tail approaches a power law with slope $p=2.4$ (dashed line). In the subplot, we show at $\ompt=1700$ the comparison between  2D and 3D results (dotted and solid line, respectively). The difference at low energies is just a consequence of the different adiabatic index between 2D and 3D. Right panel: The particle spectrum at different times is shifted along the $x$-axis by $(\ompt/15000)^{-1/2}$ and along the $y$-axis by $(\ompt/15000)^{(p-1)/2}$ with $p=2.4$, to show that the entire exponential cutoff scales in time as $\propto (\ompt)^{1/2}$. The subplot shows that the maximum Lorentz factor $\gamma_{\rm max}$ scales as $\propto (\ompt)^{1/2}$, both in 2D and in 3D (dotted and solid line, respectively).}
\label{fig:spectime0}
\end{center}
\end{figure*}

\subsection{Unmagnetized Shocks}\label{sec:pairnosig}
\begin{figure*}[tbp]
\begin{center}
\includegraphics[width=1.01\textwidth]{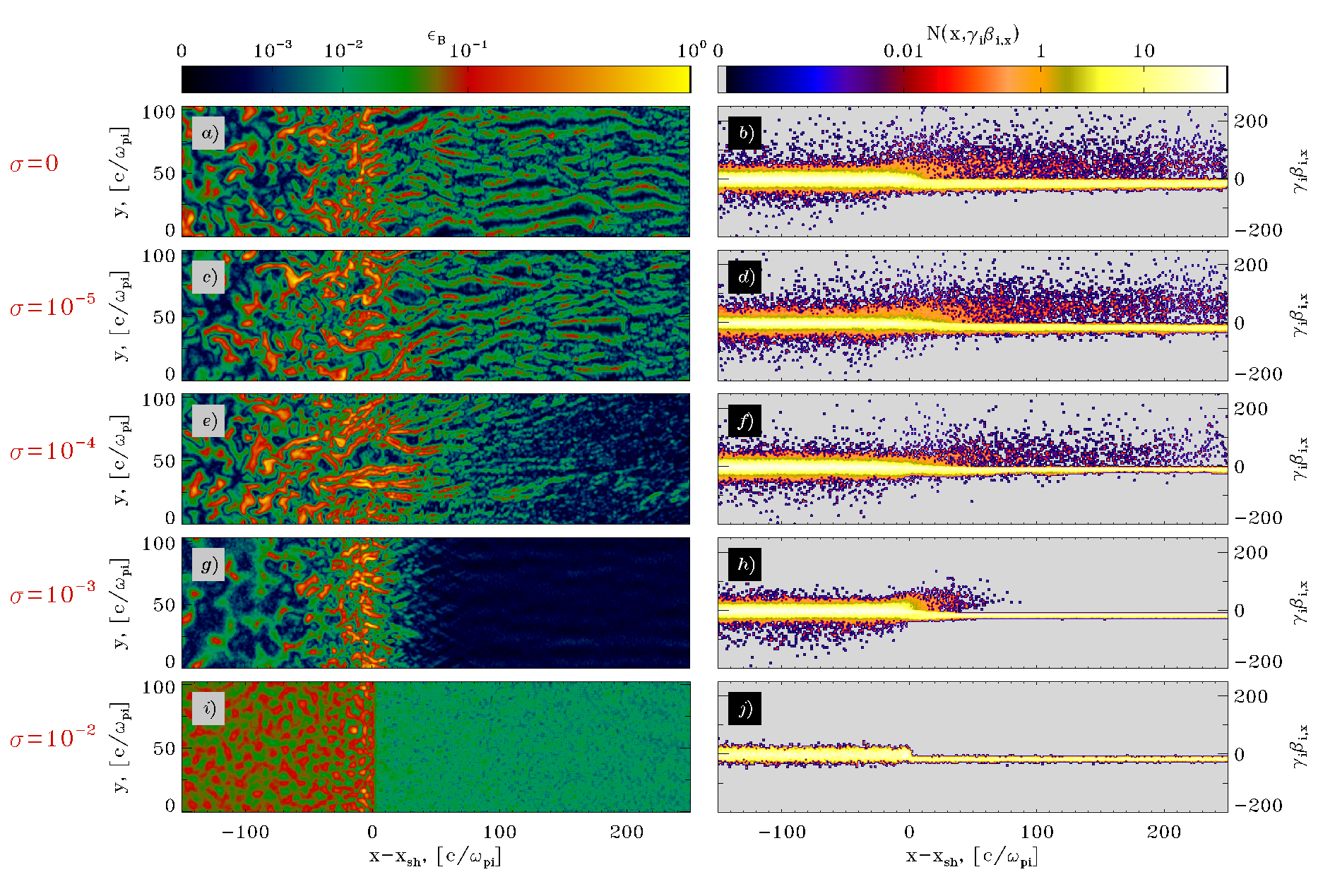}
\caption{Structure of the flow at $\ompt=4500$, from a set of 2D simulations of perpendicular electron-positron shocks with varying magnetization (from $\sigma=0$ at the top to $\sigma=\ex{2}$ at the bottom, as indicated on the left of the figure). The left column shows the 2D plot of the magnetic energy fraction $\epsilon_B=B^2/8\pi \gamma_0 n_i m_i c^2$, and the right column shows the longitudinal phase space $x-\gamma_i\beta_{i,x}$ of positrons. }
\label{fig:fluidpair}
\end{center}
\end{figure*}

As shown by \citet{spitkovsky_08b} and \citet{sironi_spitkovsky_09b} in 2D and by \citet{spitkovsky_05}  in 3D, unmagnetized shocks are mediated by the so-called Weibel instability \citep{weibel_59,medvedev_loeb_99,gruzinov_waxman_99}.\footnote{More precisely, the instability results from the coupling of the electromagnetic filamentation (Weibel) instability with the electrostatic two-stream instability, as explained by \citet{bret_09}. The resulting modes are oblique with respect to the streaming direction.} The free energy for the instability comes from the counter-streaming between the incoming flow and a beam of high-energy particles reflected back from the shock  into the upstream (``returning'' particles, from now on). The returning particles appear as a diffuse hot cloud to the right of the shock in the longitudinal phase space of \fig{fluidpair}(b), whereas the dense cold stream flowing into the shock is populated by the incoming plasma. The Weibel instability converts the kinetic energy of these two counter-propagating flows into small-scale (skin-depth) magnetic fields, organized in filaments stretching along the direction of propagation of the shock (see the 2D plot of magnetic energy in \fig{fluidpair}(a)). As the incoming flow approaches the shock, the filaments generated by the Weibel instability grow in strength and scale, eventually reaching sub-equipartition levels. At this point, they can efficiently randomize the bulk flow, and the shock forms. 

As demonstrated by  \citet{spitkovsky_08b}, unmagnetized shocks in electron-positron plasmas are efficient particle accelerators, with $\sim1\%$ of particles populating a non-thermal power-law tail of slope $p\simeq2.4$.\footnote{We define the slope $p$ of the non-thermal tail such that $dN/d\gamma\propto \gamma^{-p}$, where $\gamma$ is the particle Lorentz factor.} In \fig{spectime0}, we show the time evolution of the post-shock particle spectrum, from the 2D simulation of an electron-positron unmagnetized shock. We confirm that the non-thermal population at late times can be described as a power-law tail of slope $p\simeq2.4$, and we show that the evolution of the high-energy part of the  spectrum does not depend on the dimensionality of our computational domain. In fact, the inset of  \fig{spectime0}(a) shows that the high-energy tail is the same in 2D (dotted line) and 3D (solid line). The difference at low energies, below the peak of the thermal Maxwellian that contains most of the particles, is just a consequence of the different adiabatic index between 2D and 3D, which changes the analytic formula of a Maxwellian distribution (at low energies, we have $dN_{M\!B}/d\gamma\propto\gamma$ in 2D, as opposed to $dN_{M\!B}/d\gamma\propto\gamma\sqrt{\gamma^2-1}$ in 3D). In summary, as regards to the efficiency of particle acceleration and the shape of the power-law tail, our results show that 2D simulations provide an excellent description of the 3D physics.

As shown in \fig{spectime0}(a), the high energy cutoff of the non-thermal tail of accelerated particles shifts in time to higher and higher energies (from black to red, as time progresses). The rate of evolution of the maximum particle Lorentz factor $\gamma_{max}$ is shown in \fig{spectime0}(d), where $\gamma_{max}$ is defined as the Lorentz factor where the number of particles drops by a factor of $10^5$ with respect to the peak. Both in 2D (dotted) and in 3D (solid), the maximum Lorentz factor grows in time as $\gamma_{max}\propto (\ompt)^{1/2}$, much slower than the Bohm scaling (i.e., $\gamma_{max}\propto t$) that is assumed in most models. 

The scaling $\gamma_{max}\propto t^{1/2}$ is in agreement with analytical models of particle diffusion in strong small-scale turbulence \citep{shalchi_09,plotnikov_11,plotnikov_12}, as appropriate for the fields generated by the Weibel instability in weakly magnetized shocks. In this regime, the isotropic diffusion coefficient for an accelerated particle of energy $\varepsilon$ is 
\be\label{eq:scatt}
D(\varepsilon)\sim c \lambda\left[\frac{r_L(\varepsilon)}{\lambda}\right]^2=c \lambda\left(\frac{\varepsilon}{e B\lambda}\right)^2~~,
\ee
where $\lambda$ is the scale of the magnetic turbulence, $B$ is the mean strength of the Weibel-generated fields, and $r_L(\varepsilon)$ is the Larmor radius of a particle with energy $\varepsilon$ in the field $B$. As described by \citet{achterberg_01} and \citet{kirk_reville_10}, the fact that the spatial diffusion coefficient scales as the square of the particle energy is simply a consequence of the small-scale nature of the Weibel fluctuations. The interaction of an accelerated particle with a Weibel filament of wavelength $\lambda$ results in a deflection of the particle momentum by an angle $\sim \lambda/r_L(\varepsilon)\ll1$. The expression in \eq{scatt} simply comes from the fact that it takes $\sim [r_L(\varepsilon)/\lambda]^2$ random scatterings in the upstream turbulence to diffuse the particle direction by roughly one radian, so that the particle can cross the shock back into the downstream and continue the acceleration process. Since the  time needed to reach energy $\varepsilon$ in a relativistic shock is simply $t\sim D/c^2$ (apart from multiplicative factors of order unity), the maximum particle Lorentz factor should scale in time as
\be\label{eq:gmax}
\gamma_{max} m_i c^2\sim e B\lambda \left(\frac{ct}{\lambda}\right)^{1/2}~~.
\ee

The strength $B$ of the Weibel-generated fields can be parameterized in terms of the magnetic energy fraction $\epsilon_B\equiv B^2/8\pi\gamma_0 n_i  m_ic^2$, and the  coherence length $\lambda$ of the Weibel fluctuations  (measured in the direction transverse to the flow) can be written as a multiple $\lambda_{\comp}$ of the skin depth $\comp$, so that \eq{gmax} becomes
\be\label{eq:gmax1}
\frac{\gamma_{max}}{\gamma_0}\sim(2\epsilon_B\lambda_{\comp}\ompt )^{1/2}~~.
\ee
The scaling $\gamma_{max}\propto(\ompt)^{1/2}$ in our simulations is then expected if the combination $\epsilon_B\lambda_{\comp}$ is nearly constant in time. The temporal evolution of the magnetic structure in unmagnetized shocks has been studied by \citet{keshet_09}. They found that the characteristic transverse scale of the upstream Weibel turbulence tends to increase with time, as particles are accelerated to higher energies. However, the increase is not very significant, with $\lambda_{\comp}$ growing by a factor of $2-3$ between $\ompt\simeq10^3$ and $\ompt\simeq10^4$, corresponding to a tentative temporal scaling $\lambda_{\comp}\propto(\ompt)^{1/4}$, which is still sub-dominant with respect to the explicit temporal dependence in \eq{gmax1}. As regards to the  magnetic fraction $\epsilon_B$, \citet{keshet_09} found that  $\epsilon_B$ peaks at the shock, where its value is $\epsilon_{B,sh}\simeq0.1$ at all times. We find that the magnetic energy grows ahead of the shock with a longitudinal length scale increasing in time as $L_B\propto t$, while the particles are accelerated to higher energies. So, in a fixed volume around the shock, the average magnetic energy fraction scales as $\epsilon_{B,sh} L_B\propto t$. However, in the context of particle acceleration, one should rather compute the mean magnetic energy within a volume defined by the diffusion length of the highest energy particles, that equals $D/c$ and so scales as  $\propto t$. It follows that the mean magnetic energy fraction $\epsilon_{B}$ in a slab whose thickness is always the diffusion length of the highest energy particles is a constant fraction of $\epsilon_{B,sh}$, and so independent of time. 

In summary, the scaling $\gamma_{max}\propto(\ompt)^{1/2}$ is indeed expected in relativistic unmagnetized shocks mediated by the Weibel instability. From the inset of \fig{spectime0}(b), we can measure the coefficient of the scaling, and we obtain
\be\label{eq:gmax2}
\frac{\gamma_{max}}{\gamma_0}\simeq0.5\,(\ompt)^{1/2}~~.
\ee
It is worth pointing out that the coefficient found above depends on our definition of $\gamma_{max}$, taken here to be the Lorentz factor where the number of particles drops by a factor of $10^5$ with respect to the peak (so, where $\gamma dN/d\gamma$ is $10^5$ times lower than the peak). For instance, we obtain a different coefficient (but the same temporal scaling) if $\gamma_{max}$ is defined by using $\gamma^2 dN/d\gamma$, rather than $\gamma dN/d\gamma$. The robustness of the scaling $\gamma_{max}\propto(\ompt)^{1/2}$ is demonstrated in the right panel of \fig{spectime0}, where the particle spectrum at different times is shifted along the $x$-axis by $(\ompt/15000)^{-1/2}$ and along the $y$-axis by $(\ompt/15000)^{(p-1)/2}$ with $p=2.4$.  The fact that the exponential cutoffs of all the spectra overlap in the right panel of \fig{spectime0} suggests that the downstream non-thermal tail can be described at all times as a power law of fixed slope $p=2.4$  terminating  in an exponential cutoff that increases as  $\propto(\ompt)^{1/2}$. In other words, the scaling of the maximum energy as $\propto(\ompt)^{1/2}$ does not depend on our definition of $\gamma_{max}$, provided that $\gamma_{max}$ falls in the exponential cutoff of the particle distribution.

\begin{figure}[tbp]
\begin{center}
\includegraphics[width=0.5\textwidth]{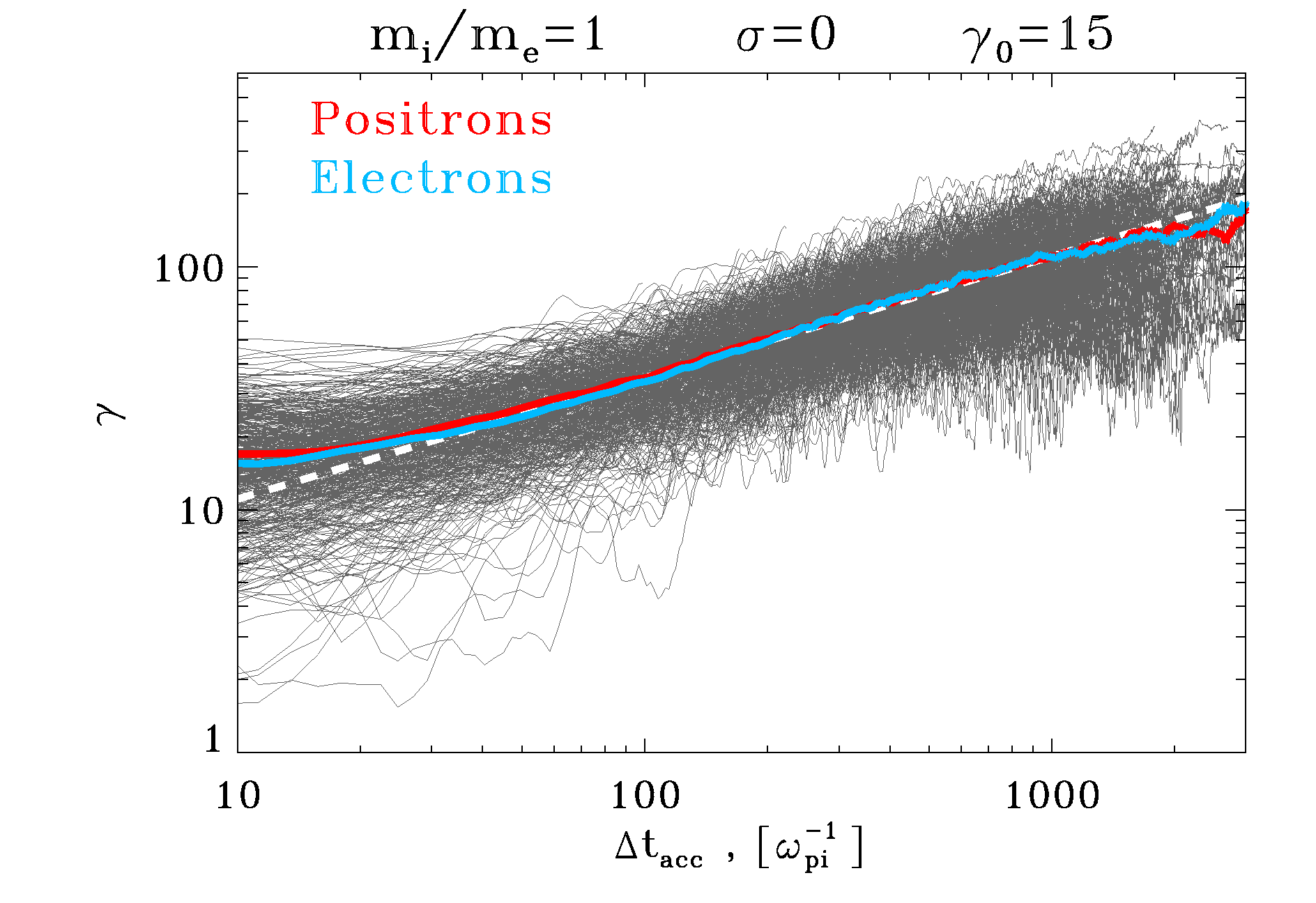}
\caption{Temporal evolution of the Lorentz factor $\gamma$ of shock-accelerated particles (as a function of the time $\Delta t_{acc}$ spent in the acceleration process), from a sample of particles extracted from the 2D simulation of an electron-positron unmagnetized shock. The particles are selected at $\ompt=4500$ such that they belong to the non-thermal tail (i.e., $\gamma\gtrsim50$ at $\ompt=4500$). Their mean Lorentz factor as a function of $\Delta t_{acc}$ is shown with the solid lines (red for positrons, blue for electrons), and it follows the $\propto \Delta t_{acc}^{1/2}$ scaling indicated by the white dashed line.}
\label{fig:testprt}
\end{center}
\end{figure}

The scaling with time in \eq{gmax2} pertains to the highest energy particles accelerated in unmagnetized electron-positron shocks. In \fig{testprt}, we show that the same scaling describes the acceleration process of all the particles belonging to the non-thermal tail. We randomly select a sample of particles that at $\ompt=4500$ have a Lorentz factor larger than $50$ (i.e., they are part of the non-thermal component at that time), and we extract their energy histories at earlier times. For each particle, we compute the Lorentz factor $\gamma$ as a function of $\Delta t_{acc}$, i.e, the time spent in the acceleration process (starting from the first crossing of the shock). \fig{testprt} shows that, for both electrons (blue) and positrons (red), the mean  energy of the accelerated particles scales as $\propto (\omega_{\rm pi}\Delta t_{acc})^{1/2}$ (compare with the white dashed line, proportional to $\Delta t_{acc}^{1/2}$). In summary, by scattering off the small-scale Weibel turbulence, the energy of the particles injected into the acceleration process grows as $\propto (\omega_{\rm pi}\Delta t_{acc})^{1/2}$. In turn, this drives the evolution $\gamma_{max}\propto (\ompt)^{1/2}$ of the upper energy cutoff in the post-shock particle spectrum.



\begin{figure*}[tbp]
\begin{center}
\includegraphics[width=1\textwidth]{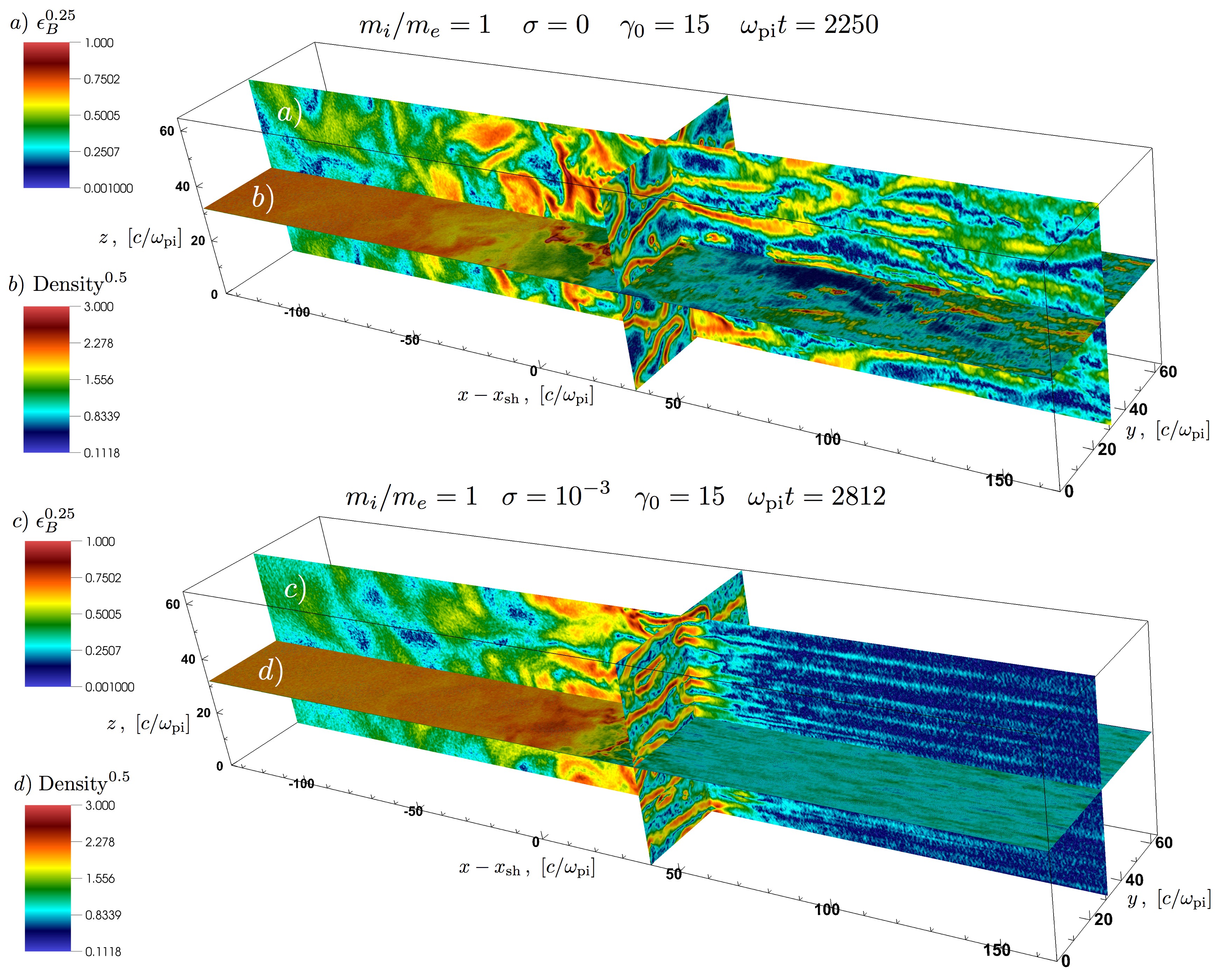}
\caption{Structure of the flow, from the 3D simulation of an electron-positron shock with  magnetization $\sigma=0$ (top) or $\sigma=\ex{3}$ (bottom). The $xy$ slice shows the particle density (with color scale stretched for clarity), whereas the  $xz$ and $yz$ slices show the magnetic energy fraction $\epsilon_B$ (with color scale stretched for clarity).}
\label{fig:fluid3d}
\end{center}
\end{figure*}

\subsection{Weakly Magnetized Shocks}\label{sec:pairsig}
In this section, we investigate the physics and acceleration properties of weakly magnetized shocks, in the regime $10^{-5}\lesssim\sigma\lesssim\ex{1}$. In \S\ref{sec:struct} we show how the structure of the shock changes when varying the magnetization, and in  \S\ref{sec:spec} we examine the dependence on $\sigma$ of the post-shock particle spectrum.

\subsubsection{Shock Structure}\label{sec:struct}
\fig{fluidpair} shows the flow structure from a set of 2D simulations of electron-positron perpendicular shocks, with magnetization increasing from $\sigma=0$ in the top row up to $\sigma=\ex{2}$ in the bottom row. We plot the magnetic energy fraction $\epsilon_B=B^2/8\pi \gamma_0 n_i m_i c^2$ in the left column (here $B$ includes both the background ordered magnetic field $B_0$ and the self-generated fields), and the positron longitudinal phase space $x-\gamma_i\beta_{i,x}$ in the right column.

With increasing magnetization from $\sigma=0$ up to $\sigma=\ex{2}$ (from top to bottom in \fig{fluidpair}), the region around the shock filled with Weibel filaments becomes narrower, and for $\sigma\gtrsim\ex{2}$ the shock structure is entirely dominated by the shock-compression of the background field $B_0$, with no evidence for Weibel-mediated turbulence. Here, the Larmor gyration of the incoming particles in the shock-compressed field is faster than the growth time of the filamentation instability, and the Weibel modes cannot grow \citep{pelletier_10}. This is in agreement with the results by SS09 and SS11, showing that the Weibel instability is suppressed in perpendicular relativistic shocks with $\sigma=\ex{1}$. 

The right column of \fig{fluidpair} shows that the longitudinal thickness of the upstream region filled with Weibel filaments is correlated with the typical distance traveled by the returning particles into the upstream (the returning particles populate the diffuse hot cloud ahead of the shock in the right column of  \fig{fluidpair}). The characteristic propagation length of the returning beam can be estimated as the Larmor radius of the returning particles in the background magnetic field $B_0$ \cite[see][]{pelletier_10}. As for the typical Lorentz factor  of the returning beam, we take the low-energy end of the non-thermal tail $\gamma_{inj}=\eta_{inj}\gamma_0$, where $\eta_{inj}\simeq5$. Since the power-law slope of the non-thermal tail is $p>2$, the low-energy end will dominate both the number and the energy census. The longitudinal thickness of the upstream region filled with Weibel turbulence will then be
\be\label{eq:foot}
L_{B,sat}\sim \frac{\eta_{inj}}{\sigma^{1/2}}\comp~~.
\ee
It is worth pointing out that the magnetic energy profile will be described by this scale $L_{B,sat}$ only at late times (see \S\ref{sec:spec} for details on the particle spectrum at late times, and the role of $L_{B,sat}$ in setting the maximum energy of the shock-accelerated particles). For earlier times, the magnetic energy  will grow in the upstream on a longitudinal scale evolving as $L_B\propto t$, as described in \S\ref{sec:pairnosig} in the special case of $\sigma=0$ shocks. Indeed, for $\sigma=0$ the expression in \eq{foot} diverges, meaning that the characteristic width of the upstream region filled with Weibel filaments will steadily increase over time.

As opposed to $L_{B,sat}$, which clearly depends on the magnetization of the flow (see the left column in \fig{fluidpair}), the magnetic energy at the shock always reaches $\sim10\%$ of the kinetic energy of the pre-shock flow, i.e., $\epsilon_{B,sh}\sim0.1$ regardless of $\sigma$. Similarly, the  transverse scale $\lambda$ of the Weibel filaments at the shock is nearly insensitive to $\sigma$, and it equals $\lambda\sim20\comp$ (left column in \fig{fluidpair}). 

As it is apparent in the magnetic energy plots of \fig{fluidpair} (left column), the Weibel modes appear in 2D  as magnetic filaments stretched in the direction of propagation of the shock. Their 3D structure is  shown in \fig{fluid3d}, for a relativistic electron-positron shock with magnetization $\sigma=0$ (top panel) and $\sigma=\ex{3}$ (bottom panel). The background magnetic field $B_0$ is oriented here along the $z$ direction, in the same way as for our 2D simulations. The $yz$ slice of the magnetic energy fraction in \fig{fluid3d}(c) shows that for $\sigma=\ex{3}$ the magnetic field ahead of the shock is primarily organized in pancakes stretched in the direction orthogonal to the background magnetic field (i.e., along  $y$). This can be simply understood, considering that the Weibel instability is seeded by the focusing of counter-streaming particles into  channels of charge and current. In the absence of a background magnetic field, the currents tend to be organized into cylindrical filaments, as demonstrated by \citet{spitkovsky_05} and shown in the $yz$ slice of the top panel in \fig{fluid3d}. In the presence of an ordered magnetic field along $z$, the particles will preferentially move along the magnetic field (rather than orthogonal), so that their currents will more likely be focused at certain locations of constant $z$, into sheets elongated along the $xy$ plane. This explains the structure of the magnetic turbulence ahead of the shock in the bottom panel of \fig{fluid3d}, common to all the cases of weakly magnetized shocks we have investigated (i.e., $0<\sigma\lesssim\ex{1}$).


\begin{figure}[ht]
\begin{center}
\includegraphics[width=0.5\textwidth]{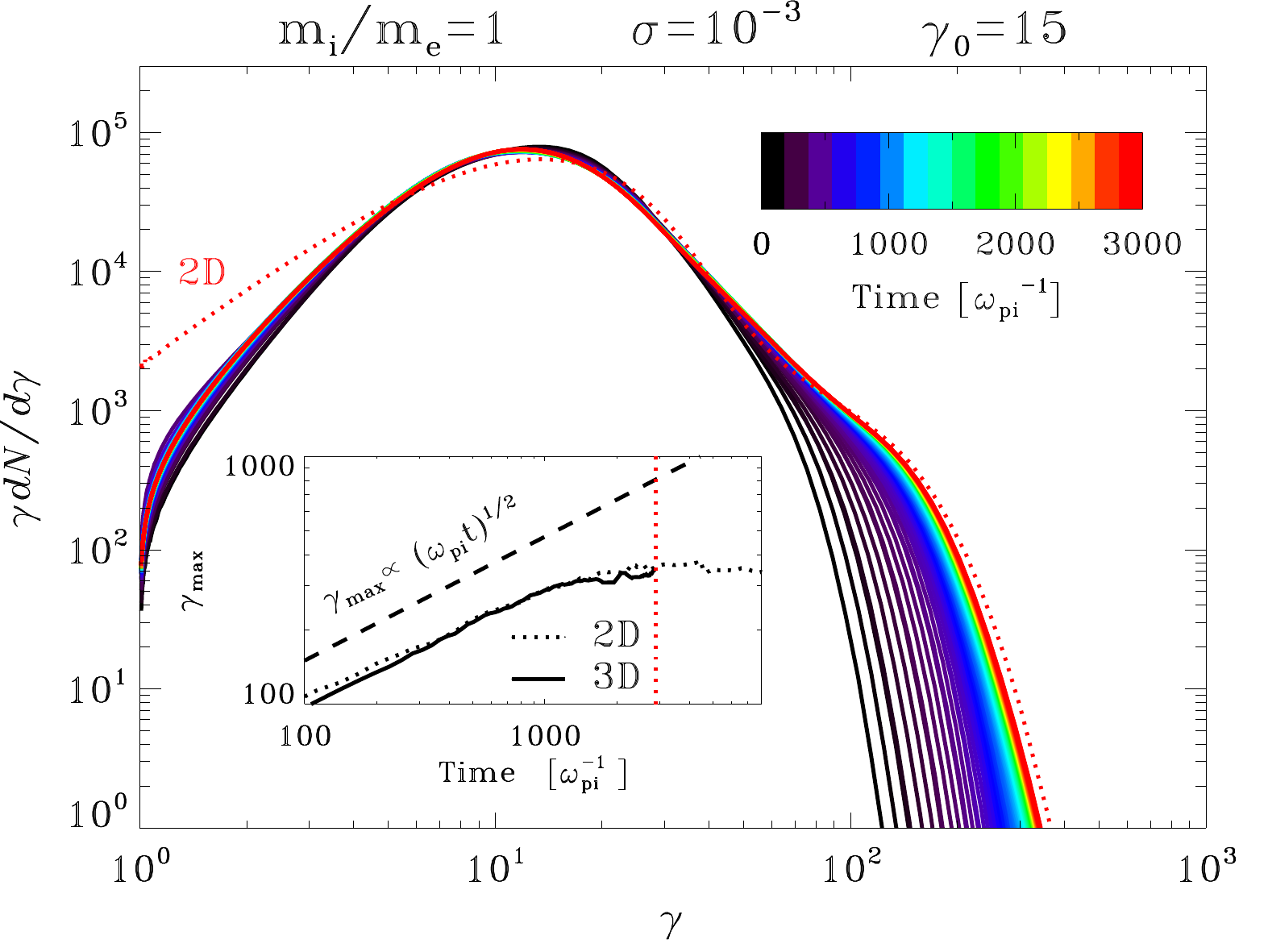}
\caption{Temporal evolution of the post-shock particle spectrum, from the 3D simulation of a $\gamma_0=15$ shock with $\sigma=\ex{3}$. We follow the evolution of the shock from its birth (black curve) up to $\ompt=3000$ (red curve). The dotted red line is the particle spectrum at $\ompt=3000$ from a 2D simulation with the same pre-shock conditions. At late times, the particle spectrum reaches a steady state, and the maximum Lorentz factor saturates, as shown in the subplot (solid line for 3D, dotted for 2D).}
\label{fig:spectimesig}
\end{center}
\end{figure}

\begin{figure*}[tb]
\begin{center}
\includegraphics[width=1\textwidth]{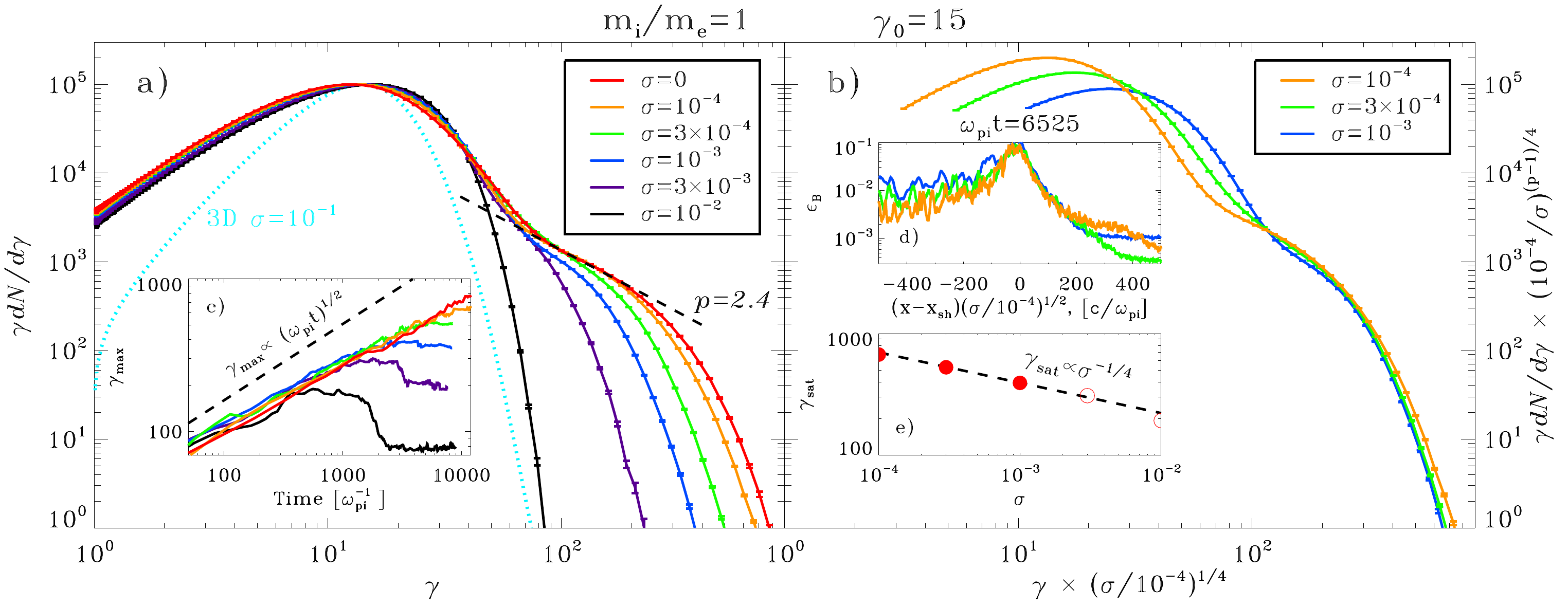}
\caption{Left panel: Dependence of the post-shock particle spectrum on the upstream magnetization, from a set of 2D simulations of electron-positron shocks with $\gamma_0=15$. We vary the magnetization between $\sigma=0$ (red curve) up to $\sigma=\ex{2}$ (black curve), showing that the Fermi process is suppressed with strong pre-shock fields. This is confirmed by the post-shock spectrum of a 3D simulation with $\sigma=\ex{1}$ (dotted cyan line). In the inset, we follow the maximum particle Lorentz factor over time, from the growth as $\gamma_{max}\propto(\ompt)^{1/2}$ to the saturation at a constant $\gamma_{sat}$. Right panel: The particle spectrum for different magnetizations is shifted along the $x$-axis by $(\sigma/10^{-4})^{1/4}$ and along the $y$-axis by $(\ex{4}/\sigma)^{(p-1)/4}$ with $p=2.4$, to show that the entire exponential cutoff scales with magnetization as  $\propto \sigma^{-1/4}$. In the inset (d), we show that the magnetic energy profile ahead of the shock has a characteristic longitudinal scale $L_{B,sat}\propto \sigma^{-1/2}$. In the inset (e), we show that the Lorentz factor at saturation $\gamma_{sat}$ scales with the magnetization as $\gamma_{sat}\propto \sigma^{-1/4}$.}
\label{fig:specsig}
\end{center}
\end{figure*}

\subsubsection{Particle Spectrum and Acceleration}\label{sec:spec}
We now explore the acceleration performance of weakly magnetized electron-positron shocks. In \fig{spectimesig} we follow the evolution of the post-shock particle spectrum from  the 3D simulation of a shock with magnetization $\sigma=\ex{3}$. In \fig{specsig} we compare the particle energy spectra at late times for different magnetizations, covering the range $0\lesssim\sigma\lesssim\ex{1}$.

As compared to the results for unmagnetized shocks in \fig{spectime0}, the evolution of the post-shock energy spectrum in \fig{spectimesig} for a flow with $\sigma=\ex{3}$ shows that the non-thermal tail initially grows to higher energies, but then it saturates (all the curves for $\ompt\gtrsim1500$ overlap). The saturation of the high-energy tail is a robust result, holding in 3D (solid lines) and in 2D (dotted red line at $\ompt=3000$), and it is clearly in contrast with the steady increase of the high-energy spectral cutoff observed for unmagnetized shocks in  \fig{spectime0}. The inset in \fig{spectimesig} confirms that the maximum Lorentz factor initially increases as $\gamma_{max}\propto(\ompt)^{1/2}$ (solid line for 3D, dotted for 2D), similarly to the case of unmagnetized shocks, but for $\ompt\gtrsim1500$ it saturates at $\gamma_{sat}\simeq350$.

We find that the scaling $\gamma_{max}\propto(\ompt)^{1/2}$, followed by saturation at a constant $\gamma_{sat}$, is a common by-product of the evolution of all magnetized relativistic shocks. In \fig{specsig}(a) we show several post-shock spectra for different magnetizations, after the non-thermal tail has reached the saturation stage. We cover the range $10^{-4}\lesssim\sigma\lesssim\ex{1}$, and for the sake of completeness we compare our results with the unmagnetized case $\sigma=0$, where the non-thermal tail is still evolving to higher and higher energies. We find that strongly magnetized electron-positron shocks, with $\sigma\gtrsim\ex{2}$, are poor particle accelerators, in agreement with the conclusions of SS09. The post-shock spectrum at late times (see the black solid line for $\sigma=\ex{2}$) is fully consistent with a Maxwellian distribution. This result does not depend on the reduced dimensionality of our 2D computational domain. We have performed a large-scale 3D simulation of an electron-positron perpendicular shock with $\sigma=\ex{1}$, and we confirm that the post-shock particle spectrum (dotted cyan line in \fig{specsig}(a)) does not show any evidence for particle acceleration. This undoubtedly proves that the absence of accelerated particles in the 2D simulations of perpendicular strongly magnetized shocks performed by SS09 is a physical consequence of the lack of sufficient self-generated turbulence,\footnote{More precisely, the fluctuations that get self-excited in $\sigma\gtrsim\ex{2}$ shocks (cyclotron modes and their harmonics) have a short path length for emission and absorption, so they constantly enforce the local thermal equilibrium, giving Maxwellian energy spectra.} rather than an artifact of the reduced dimensionality of the simulation box, as argued by \citet{jones_98}.

For weakly magnetized shocks, with $\sigma\lesssim3\times\ex{3}$, we find efficient particle acceleration, with a non-thermal tail of slope $p\simeq 2.4$ (dashed black line in \fig{specsig}(a))) that contains $\sim1\%$ of particles and $\sim10\%$ of flow energy, regardless of the magnetization. The low-energy end of the non-thermal tail does not significantly depend on the magnetization ($\gamma_{inj}\simeq5\gamma_0$, or equivalently $\eta_{inj}\simeq5$), but the high-energy cutoff at saturation is systematically higher for lower magnetizations. This is confirmed by the inset of \fig{specsig}(a), where we plot the evolution in time of the maximum Lorentz factor $\gamma_{max}$. Regardless of the magnetization, $\gamma_{max}$ initially grows as $\gamma_{max}\propto(\ompt)^{1/2}$, with a coefficient of proportionality that does not significantly depend on $\sigma$. At later times, the maximum energy departs from this scaling, and it saturates at a Lorentz factor $\gamma_{sat}$ which is larger for smaller magnetizations. 

For relatively high magnetizations (black for $\sigma=\ex{2}$ and purple for $\sigma=3\times\ex{3}$ in the inset of \fig{specsig}(a)), the maximum energy initially grows as $\propto(\ompt)^{1/2}$, then  saturates at $\gamma_{sat}$, and finally  drops to a smaller value. The drop at late times is driven by a decrease in the strength of the Weibel turbulence, as a result of heating in the pre-shock medium \citep[see][for a discussion of temperature effects on the growth of beam-plasma instabilities]{bret_10b}. The pre-shock heating is induced by the so-called electromagnetic precursor wave, a train of transverse electromagnetic waves propagating into the upstream as a result of the synchrotron maser instability at the shock front \citep{hoshino_91, hoshino_08}. The electromagnetic precursor heats the incoming particles, thus suppressing the Weibel instability and the particle acceleration at late times. Since the strength of the precursor wave decreases for smaller magnetizations \citep{gallant_92}, for $\sigma\lesssim\ex{3}$ the pre-heating induced by the electromagnetic precursor does not appreciably affect the structure of the magnetic turbulence and the process of particle acceleration. So, for $\sigma\lesssim\ex{3}$ the maximum Lorentz factor does not drop at late times.

The scaling of the Lorentz factor at saturation $\gamma_{sat}$ with respect to $\sigma$ is studied in the right panel of \fig{specsig}. In \fig{specsig}(d) we demonstrate that the magnetic energy profile ahead of the shock is characterized by a longitudinal scale $L_{B,sat}\propto\sigma^{-1/2}$, as suggested in \eq{foot}. The curves in  \fig{specsig}(d) refer to $\ompt=6525$, when the shock has already reached a steady state, and the maximum energy of the non-thermal tail has saturated (see \fig{specsig}(c)). Since the magnetic turbulence that governs the  acceleration process only extends across a region of longitudinal thickness $L_{B,sat}$ ahead of the shock, the diffusion length of the highest energy particles will be limited by the requirement $D/c\lesssim L_{B,sat}$.\footnote{In reality, the Fermi-accelerated particles will sample both the upstream and the downstream turbulence, whereas $L_{B,sat}$ only pertains to the upstream side of the shock. Since the decay length of the Weibel turbulence in the downstream is still a matter of debate \citep{chang_08,keshet_09}, here we take $L_{B,sat}$ as our best estimate for the overall thickness of the turbulent region around the shock, including both  upstream and  downstream.} The Fermi process cannot proceed to higher energies, due to the lack of sufficient magnetic turbulence on the relevant diffusive scales. More precisely,  in Weibel-mediated shocks the non-thermal particles need to be continuously grazing the shock surface in order to be further accelerated, since the wavelength of the Weibel modes is too small to sustain large-angle scatterings \citep[as described by][]{spitkovsky_08b}. If a hypothetical particle were to penetrate into the upstream beyond the longitudinal scale $L_{B,sat}$ of the Weibel turbulence, it would be deflected by the ordered background field away from its grazing trajectory into the downstream. In the absence of sufficient downstream turbulence, it will leave the shock transition region, preventing further acceleration. 
The requirement $D/c\lesssim L_{B,sat}$ sets the particle Lorentz factor at saturation to be
\be\label{eq:gsat1}
\frac{\gamma_{sat}}{\gamma_0}\sim\left(\frac{2\epsilon_B\lambda_{\comp}\eta_{inj}}{\sqrt{\sigma}}\right)^{1/2}~~,
\ee
which implies, since $\epsilon_B$, $\lambda_{\comp}$ and $\eta_{inj}$ do not significantly depend on $\sigma$, that $\gamma_{sat}/\gamma_0\propto \sigma^{-1/4}$. This scaling is demonstrated in \fig{specsig}(e),\footnote{In \fig{specsig}(e), the open circles for $\sigma=3\times\ex{3}$ and $\sigma=\ex{2}$ show the Lorentz factor that is reached during the plateau phase after the initial growth as $\propto(\ompt)^{1/2}$, and before the drop at late times induced by the electromagnetic procursor (see the black and purple lines in \fig{specsig}(c)).} where we measure the coefficient of proportionality to be
\be\label{eq:gsat2}
\frac{\gamma_{sat}}{\gamma_0}\simeq\frac{4}{\sigma^{1/4}}~~.
\ee
Finally, the right panel of \fig{specsig} (yellow curve for $\sigma=\ex{4}$, green for $3\times\ex{4}$ and blue for $\ex{3}$), where we rescale the $x$-axis by $(\sigma/10^{-4})^{1/4}$ and the $y$-axis by $(10^{-4}/\sigma)^{(p-1)/4}$ with $p=2.4$, demonstrates that  the entire exponential cutoff of the non-thermal tail scales as $\propto \sigma^{-1/4}$, and so the proportionality $\gamma_{sat}/\gamma_0\propto \sigma^{-1/4}$ does not depend on the details of our definition of $\gamma_{sat}$ (yet, the coefficient of the scaling in \eq{gsat2} does depend on the definition of $\gamma_{sat}$).


\subsection{Dependence on the Lorentz Factor $\gamma_0$}\label{sec:gamma}
\begin{figure}[tbp]
\begin{center}
\includegraphics[width=0.5\textwidth]{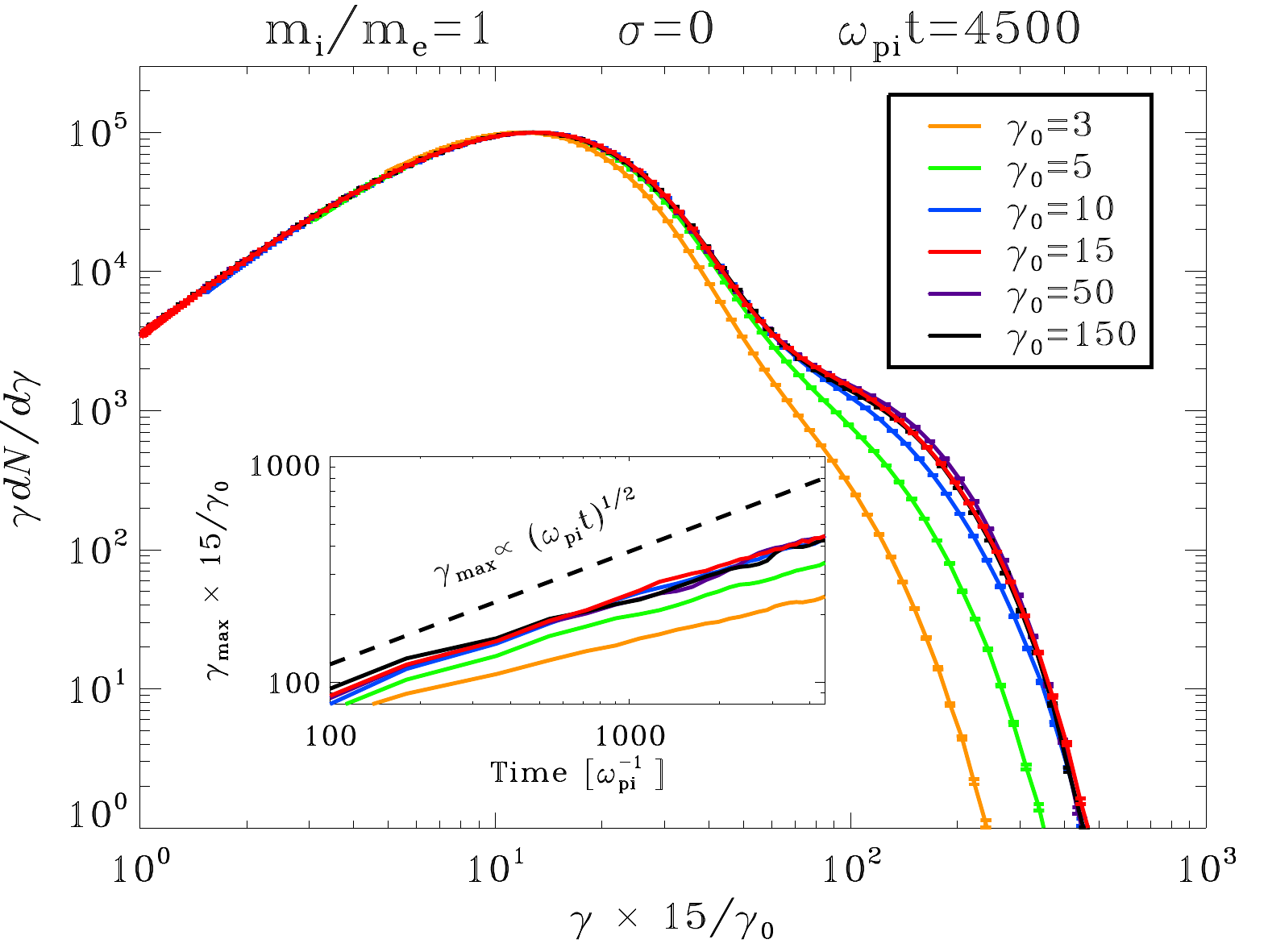}
\caption{Dependence of the post-shock particle spectrum on the upstream bulk Lorentz factor $\gamma_0$, from a set of 2D simulations of unmagnetized electron-positron shocks. The spectra are shifted along the $x$-axis by $15/\gamma_0$, to facilitate comparison with the reference case $\gamma_0=15$. In the subplot, we show that the maximum Lorentz factor $\gamma_{\rm max}$ scales as $\propto (\ompt)^{1/2}$ for all the values of $\gamma_0$ we explore. Both the main plot and the subplot show that the physics of relativistic shocks does not depend on $\gamma_0$, for $\gamma_0\gtrsim10$.}
\label{fig:specgam0}
\end{center}
\end{figure}

\begin{figure}[tbp]
\begin{center}
\includegraphics[width=0.5\textwidth]{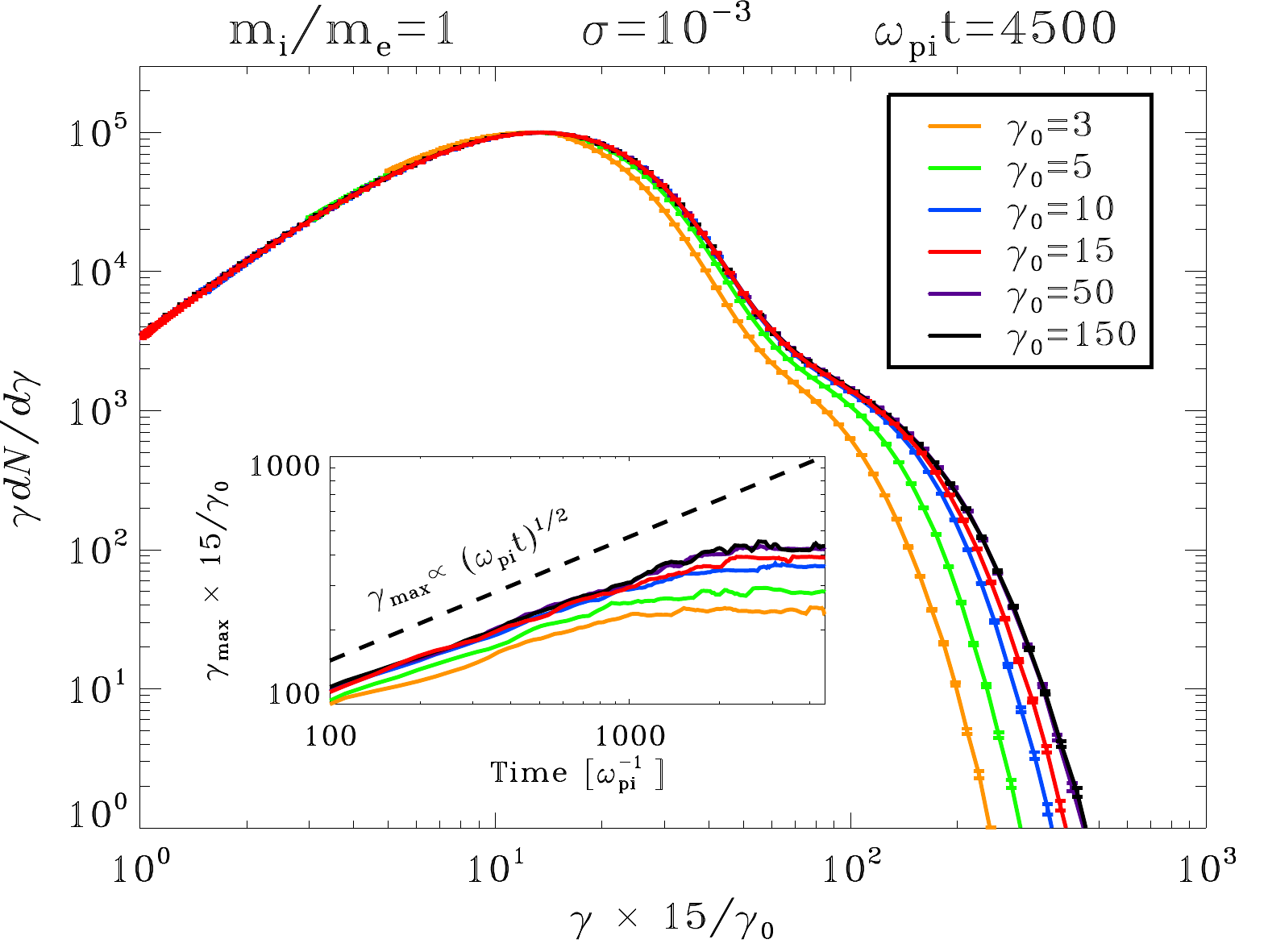}
\caption{Dependence of the post-shock particle spectrum on the upstream bulk Lorentz factor $\gamma_0$, from a set of 2D simulations of $\sigma=\ex{3}$ electron-positron shocks. See the caption of \fig{specgam0} for further details. For all the values of $\gamma_0$, the maximum particle energy saturates after $\ompt\simeq1500$.}
\label{fig:specgamsig}
\end{center}
\end{figure}

In Figs.~\ref{fig:specgam0} and \ref{fig:specgamsig}, we show the dependence of the post-shock particle spectrum on the upstream bulk Lorentz factor $\gamma_0$, for both unmagnetized (\fig{specgam0}) and weakly magnetized ($\sigma=\ex{3}$; \fig{specgamsig}) electron-positron shocks. We vary $\gamma_0$ from $\gamma_0=3$ (a mildly relativistic shock) up to $\gamma_0=150$ (an ultra-relativistic shock).

For both unmagnetized and weakly magnetized shocks, we find that the acceleration physics does not depend on the bulk Lorentz factor of the upstream flow, if $\gamma_0\gtrsim10$. Both the acceleration efficiency and the rate of particle energization are insensitive to $\gamma_0$, in the limit of ultra-relativistic flows. In particular, the insets in Figs.~\ref{fig:specgam0} and \ref{fig:specgamsig} show that the temporal evolution of $\gamma_{max}/\gamma_0$ is the same for all $\gamma_0\gtrsim10$, as predicted by eqs.~(\ref{eq:gmax2}) and (\ref{eq:gsat2}). As shown in the inset of \fig{specgamsig}, for $\sigma=\ex{3}$ both the scaling $\gamma_{max}\propto (\ompt)^{1/2}$ at early times and the saturation of the maximum energy at late times hold regardless of $\gamma_0\gtrsim10$.

The decrease in the acceleration efficiency for $\gamma_0\lesssim10$ is a consequence of the fact that the Weibel and oblique instabilities are suppressed when the beam of returning particles has a significant transverse dispersion (i.e., it is not cold, in the pre-shock frame). The instability will be quenched if $\omega_g\lesssim \bmath{k}\cdot\bmath{\Delta v}$, where $\omega_g$ is the growth rate of the most unstable mode, that usually occurs for $k\sim k_{\perp}\sim \omega_{\rm pi}/c$.\footnote{Here we only consider electron-positron plasmas, so $\comp=c/\omega_{\rm pe}$. For electron-ion flows, see \citet{pelletier_11}.} In the post-shock frame, the parallel and perpendicular components of the momentum of the returning beam are $p_{\parallel}\sim p_\perp\sim\gamma_{inj}m_i c$. In the pre-shock frame, $p'_{\parallel}\sim \gamma_0\,p_{\parallel}$ and $p'_{\perp}\sim p_{\perp}$, so that the characteristic beam transverse velocity in the pre-shock frame is $\Delta v'_{\perp}\sim c\, p'_{\perp}/p'_{\parallel}\sim c/\gamma_0$, which is larger for smaller $\gamma_0$. The beam transverse dispersion should be compared with the growth rate of the relevant instability \citep[e.g.,][]{bret_10}, which in the upstream frame of electron-positron shocks is $\omega_g\sim(\xi_b/\eta_{inj})^{1/2}\omega_{\rm pi}$ for the Weibel mode and $\omega_g\sim(\xi_b/\eta_{inj})^{1/3}\omega_{\rm pi}$ for the oblique mode. Here, $\xi_b$ is the density ratio between the beam of returning particles and the incoming flow, measured in the downstream frame, whereas $\eta_{inj}\equiv \gamma_{inj}/\gamma_0$. Both $\xi_b$ and $\eta_{inj}$ are nearly insensitive to $\gamma_0$ (Figs.~\ref{fig:specgam0} and \ref{fig:specgamsig} show that the non-thermal tails for different $\gamma_0$ are all starting at the same $\gamma_{inj}/\gamma_0$, and with similar normalizations), so that the generation of Weibel and oblique modes in electron-positron flows should be suppressed for
\be
\gamma_0\lesssim (\xi_b/\eta_{inj})^{-1/2}\;\;[{\rm Weibel}]\\
\gamma_0\lesssim  (\xi_b/\eta_{inj})^{-1/3}\;\;[{\rm oblique}]
\ee
In turn, the suppression of the Weibel and oblique instabilities, which are responsible for the generation of the magnetic turbulence that governs the Fermi process, results in the poorer acceleration capabilities of  mildly relativistic shocks, as compared to their ultra-relativistic counterparts (see the green and yellow curves in Figs.~\ref{fig:specgam0} and \ref{fig:specgamsig}, for $\gamma_0=5$ and $\gamma_0=3$ respectively).

\vspace{0.2in}
\section{Electron-Ion Shocks}\label{sec:ions}
In this section, we investigate the physics of electron-ion shocks, by employing a reduced mass ratio $\mime=25$ or $\mime=100$. As we describe in Appendix \ref{sec:specmime}, our conclusions do not change for higher mass ratios, up to the realistic value $\mime\simeq1836$. In \S\ref{sec:structmi} we show  how  the structure of the flow changes for different magnetizations, in the regime $0\lesssim\sigma\lesssim\ex{3}$. In  \S\ref{sec:specmi} we follow the time evolution of the post-shock spectrum, and we comment on the effect of the flow magnetization  on the long term evolution of the acceleration physics. Finally, in  \S\ref{sec:specgammi} we explore the dependence of our results on the bulk Lorentz factor of the pre-shock flow.

\begin{figure*}[tbp]
\begin{center}
\includegraphics[width=1\textwidth]{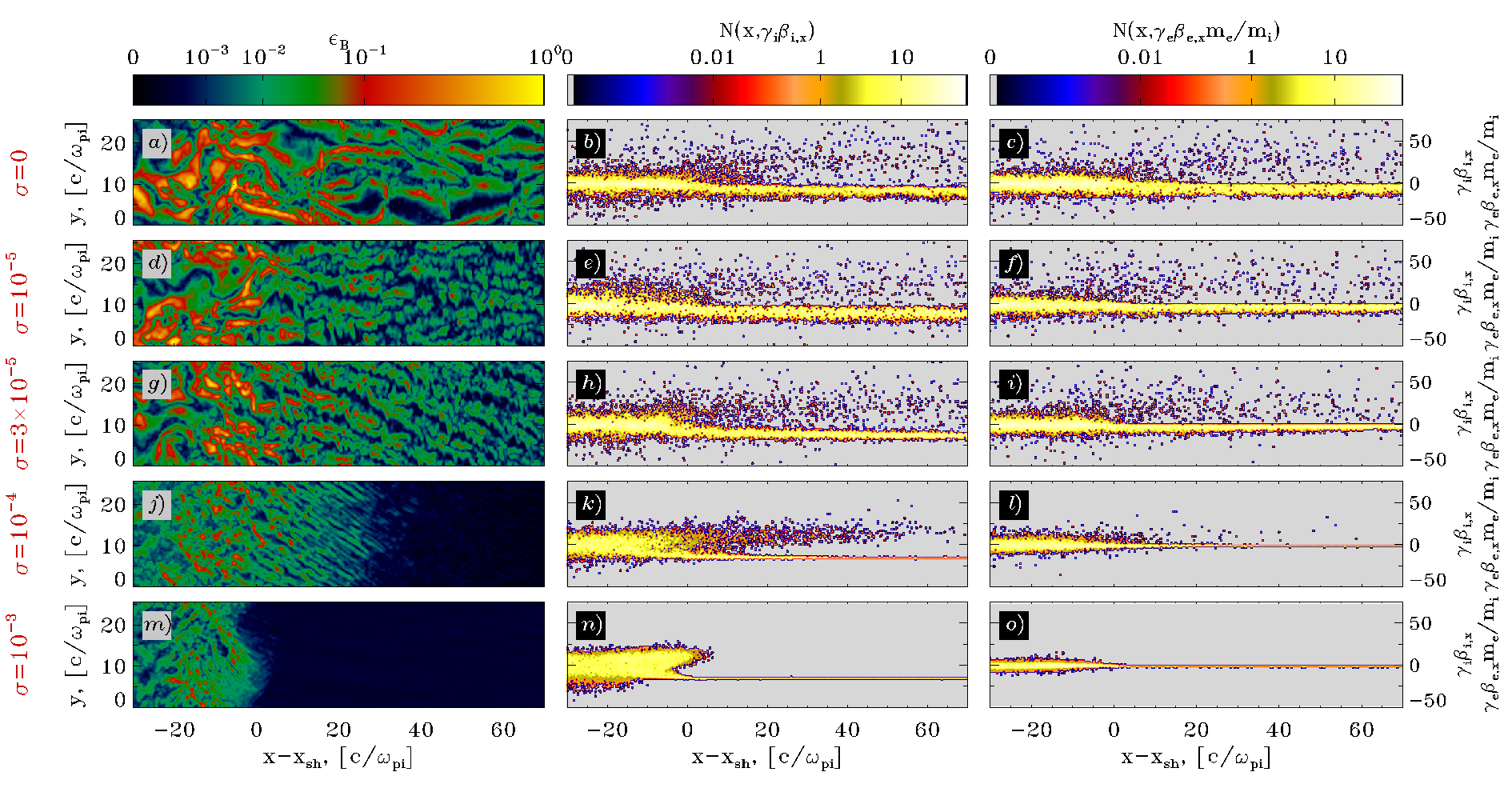}
\caption{Structure of the flow at $\ompt=2250$, from a set of 2D simulations of perpendicular electron-ion shocks ($\mime=25$) with varying magnetization (from $\sigma=0$ at the top to $\sigma=\ex{3}$ at the bottom, as indicated on the left of the figure). The first column shows the 2D plot of the magnetic energy fraction $\epsilon_B=B^2/8\pi \gamma_0 n_i m_i c^2$, the second column the longitudinal phase space $x-\gamma_i\beta_{i,x}$ of ions, and the third column the longitudinal phase space $x-\gamma_e\beta_{e,x}$ of electrons.}
\label{fig:fluidmi}
\end{center}
\end{figure*}

\subsection{Shock Structure}\label{sec:structmi}
The physics of electron-ion unmagnetized shocks has been investigated by \citet{spitkovsky_08} and \citet{martins_09} in 2D and by \citet{haugbolle_10} in 3D. They have found that electron-ion unmagnetized (i.e., $\sigma=0$) shocks are mediated by the Weibel instability, in the same way as electron-positron shocks.  The instability is seeded by the counter-streaming between the incoming flow and the ions and electrons reflected back from the shock into the upstream (see the diffuse hot cloud of ions and electrons moving ahead of the shock in \fig{fluidmi}(b) and (c), respectively). As they approach the shock, the incoming electrons are heated  by the time-varying Weibel fields \citep{gedalin_12,plotnikov_12}, and they reach equipartition with the ions before entering the shock. As the electrons get heated on their way to the shock, the typical  transverse scale of the Weibel filaments  increases from the electron skin depth  $c/\omega_{\rm pe}$ far ahead of the shock  to the proton skin depth  $c/\omega_{\rm pi}$ just in front of the shock. In the 2D plot of the magnetic energy in \fig{fluidmi}(a), the transverse scale of the Weibel filaments at the shock amounts to a few ion skin depths.

In \fig{fluidmi}, we show how the 2D structure of the magnetic energy (first column) and the longitudinal phase spaces of ions and electrons (second and third columns, respectively) change for increasing magnetization, as compared to the unmagnetized case (top row) most commonly discussed in the literature. We investigate the range $0\lesssim\sigma\lesssim\ex{3}$, from top to bottom in \fig{fluidmi}.

Similarly to the case  $\sigma=0$, in the regime $\sigma\lesssim\ex{4}$ of weakly magnetized shocks the flow structure is dominated by the Weibel  modes. In the same way as for electron-positron magnetized flows in \S\ref{sec:struct}, the thickness of the upstream region that is filled with Weibel filaments is sensitive to the flow magnetization, being set by the typical Larmor radius of returning ions, that scales as $\propto \sigma^{-1/2}$ (see \eq{foot}). The two main differences with respect to electron-positron shocks are in the transverse scale of the Weibel filaments and in their orientation relative to the shock normal, as we now discuss. 

As the magnetization increases in the range $\ex{5}\lesssim\sigma\lesssim\ex{4}$, the Weibel filaments ahead of the shock appear narrower. As anticipated above, if    the characteristic electron thermal Lorentz factor in the pre-shock frame is $\gamma_{th,e}$, the fastest growing Weibel modes will operate on a transverse scale $\sim\sqrt{\gamma_{th,e}}\compe$. For $\sigma\lesssim\ex{5}$, the region filled with Weibel filaments is wide enough to drive efficient ion-to-electron energy transfer ahead of the shock, and $\gamma_{th,e}\sim\mime$ as the electrons enter the shock. It follows that the characteristic transverse scale of the Weibel filaments in \fig{fluidmi}(d) is comparable to the ion skin depth $\comp=\sqrt{\mime}\compe$ for $\sigma\lesssim\ex{5}$. On the other hand, for $3\times\ex{5}\lesssim\sigma\lesssim\ex{4}$, the electrons ahead of the shock do not have enough time to reach equipartition with the ions, in the Weibel turbulence generated ahead of the shock. So, $\gamma_{th,e}$ stays smaller than $\mime$, and the Weibel filaments are narrower.

Since for $3\times\ex{5}\lesssim\sigma\lesssim\ex{4}$ the electrons enter the shock with energy lower than the ions (resulting in a lower dispersion of their post-shock longitudinal phase space, compare panels (k) and (l) for $\sigma=\ex{4}$), a smaller fraction of electrons will be injected into the acceleration process, relative to ions. This is demonstrated in the longitudinal phase spaces of $\sigma=3\times\ex{5}$ (panel (h) for ions and (i) for electrons) and $\sigma=\ex{4}$ (panels (k) and (l)), that show a smaller number of returning  electrons, with respect to ions. By comparison, in the cases where the ion-to-electron energy transfer ahead of the shock is efficient (i.e., $\sigma\lesssim\ex{5}$), returning electrons and ions have comparable number density and energy. If ions outnumber electrons,  as it is the case for $3\times\ex{5}\lesssim\sigma\lesssim\ex{4}$, the orientation of the Weibel modes ahead of the shock will be determined by the direction of the electric current seeded by the returning ions. For a background magnetic field in the $+\bmath{\hat{z}}$ direction, as we employ here, the ions back-streaming ahead of the shock will preferentially move with $p_y\sim - p_x<0$. The electric current  they generate is then in the same direction as the Weibel filaments seen in the upstream region of  $\sigma=3\times\ex{5}$ (panel (g)) and  $\sigma=\ex{4}$ (panel (j)) shocks. In contrast, if the injection of ions and electrons into the acceleration process is equally efficient (so that the two species equally contribute to the population of returning particles), the mean flux of returning particles stays aligned with the shock normal along $+\bmath{\hat{x}}$, and the Weibel filaments are stretched along the shock direction of propagation, as shown in panel (a) for $\sigma=0$ and panel (d) for $\sigma=\ex{5}$.

For higher magnetizations ($\sigma=\ex{3}$ in the bottom row of \fig{fluidmi}), the returning ions are confined closer to the shock by the background magnetic field (panel (n)), no returning electrons are present (panel (o)), and the structure of the shock is no longer dominated by the Weibel filaments (panel (m)). The 2D plot of magnetic energy in \fig{fluidmi}(m) shows that the shock surface is rippled, on a scale comparable to the ion Larmor radius. The phenomenon of shock rippling is well known in non-relativistic supercritical shocks \citep{burgess_06}, where it is associated with the presence of gyrating reflected ions in the shock transition layer. In our simulations of relativistic perpendicular shocks, we observe the shock ripples only in a limited regime of magnetizations, $3\times\ex{4}\lesssim\sigma\lesssim\ex{1}$. For both smaller and higher magnetizations, the rippling is suppressed by the pre-shock heating of the incoming electrons. The cause for the electron heating varies with the flow magnetization (the time-dependent Weibel fields for $3\times \sigma\lesssim\ex{4}$; the electromagnetic precursor wave for $\sigma\gtrsim\ex{1}$, as discussed by SS11), but in all the cases it results in a suppression of the shock ripples. As we discuss in \S\ref{sec:specmi}, we do not find that shock rippling facilitates the injection of particles into the acceleration process. This result has been confirmed with a simulation box three times as large as in \fig{fluidmi}(m), such that to accommodate three shock  ripples in the computational domain.


\begin{figure}[tbp]
\begin{center}
\includegraphics[width=0.5\textwidth]{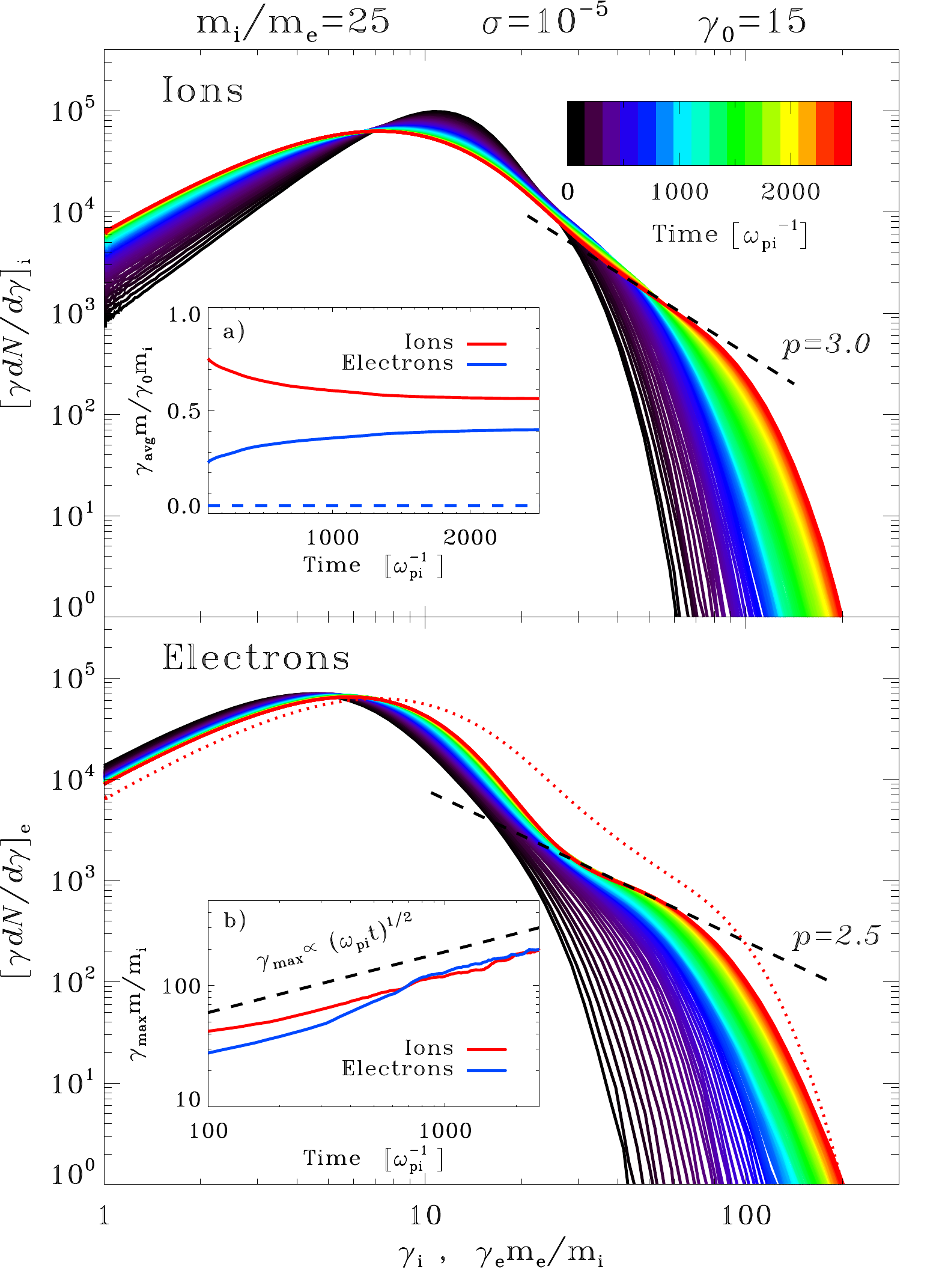}
\caption{Temporal evolution of the post-shock particle spectrum, from the 2D simulation of a $\gamma_0=15$ electron-ion ($\mime=25$) shock in a flow with magnetization $\sigma=\ex{5}$. We follow the evolution of the shock from its birth (black curve) up to $\ompt=2500$ (red curve). In the top panel we show the ion spectrum, in the bottom panel the electron spectrum. The ion non-thermal tail approaches at late times a power law with slope $p=3.0$, whereas $p=2.5$ for electrons (dashed lines). In the bottom panel, we overplot the ion spectrum at $\ompt=2500$ with a red dotted line. Inset (a): mean post-shock ion (red) and electron (blue) energy, in units of the bulk energy of the upstream flow. The dashed blue line shows the electron energy at injection. Inset (b): temporal evolution of the maximum Lorentz factor of ions (red) and electrons (blue), scaling as $\propto (\ompt)^{1/2}$ at late times (black dashed line).}
\label{fig:spectimemi}
\end{center}
\end{figure}

\begin{figure}[tbp]
\begin{center}
\includegraphics[width=0.5\textwidth]{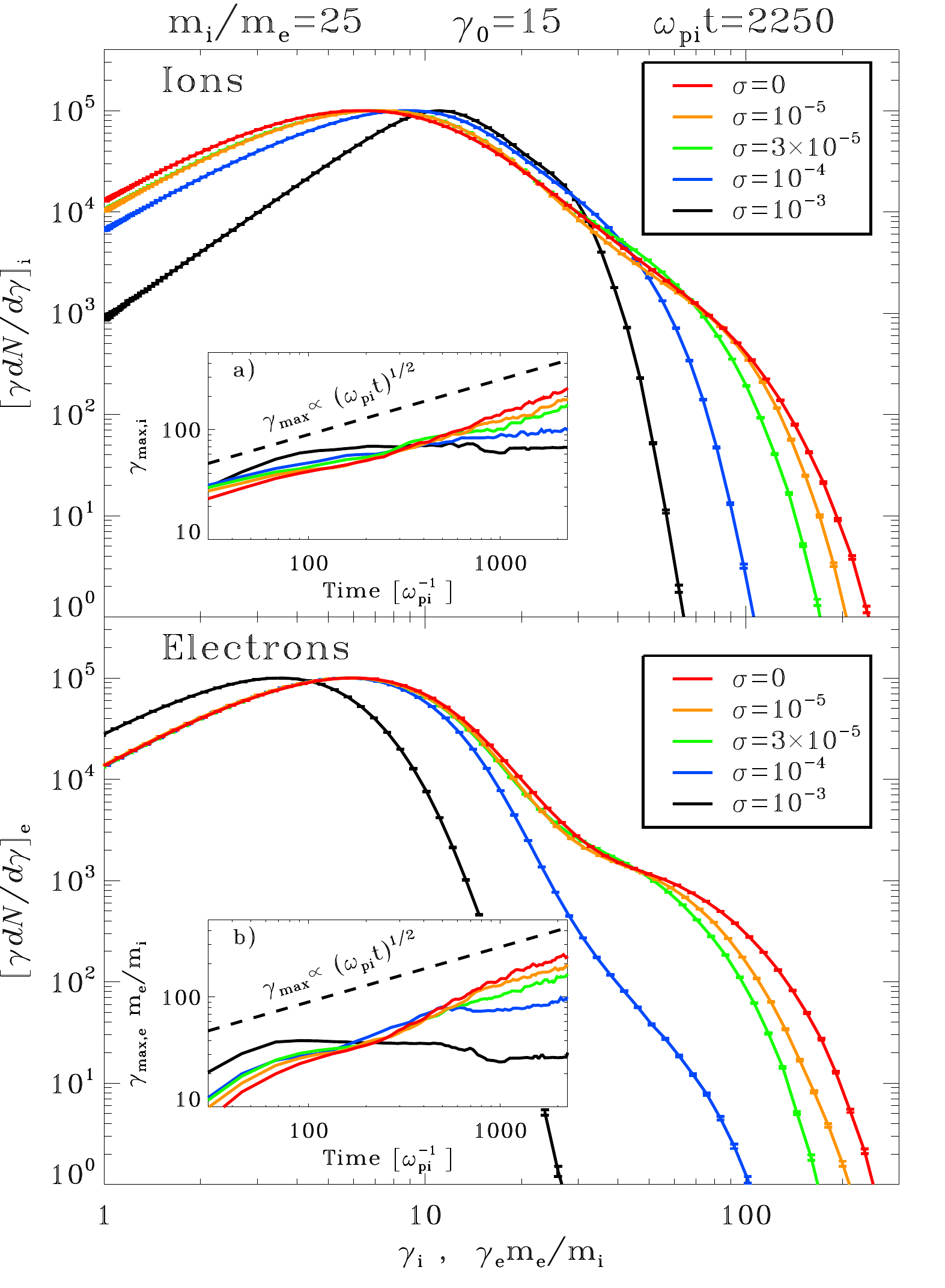}
\caption{Dependence of the post-shock particle spectrum on the upstream magnetization, from a set of 2D simulations of electron-ion shocks ($\mime=25$) with $\gamma_0=15$. In the top panel we show the ion spectrum, in the bottom panel the electron spectrum.  We vary the magnetization between $\sigma=0$ (red curve) up to $\sigma=\ex{3}$ (black curve), showing that the Fermi process is suppressed for large $\sigma$. In the insets, we follow the maximum particle Lorentz factor over time, for ions (panel (a)) and electrons (panel (b)). }
\label{fig:specsigmi}
\end{center}
\end{figure}

\subsection{Particle Spectrum and Acceleration}\label{sec:specmi}
We now investigate the  acceleration physics in perpendicular electron-ion magnetized shocks, by employing a reduced mass ratio $\mime=25$. As we show in Appendix \ref{sec:specmime}, our conclusions do not change for higher mass ratios, up to the realistic value $\mime\simeq1836$.

In \fig{spectimemi} we follow the evolution of the post-shock particle spectrum from the 2D simulation of a perpendicular shock propagating in a weakly magnetized plasma, with magnetization $\sigma=\ex{5}$. Similarly to the case of unmagnetized electron-ion shocks (i.e., $\sigma=0$) investigated by \citet{spitkovsky_08} and \citet{martins_09} in 2D and by \citet{haugbolle_10} in 3D, both the ion and the electron spectra show a prominent non-thermal power-law tail of accelerated particles, beyond the thermal distribution. 

As a result of efficient energy transfer in front of the shock, mediated by the Weibel turbulence, the electrons enter the shock nearly in equipartition with the ions. This is demonstrated in the inset (a) of \fig{spectimemi}, where we follow the temporal evolution of the mean electron and ion energies in the post-shock flow, in units of the initial ion energy. At late times, the mean electron energy (blue line) is a fraction $\sim40\%$ of the initial ion bulk energy, and it amounts to a fraction $\sim65\%$ of the mean post-shock ion energy (red line). This is much larger than the initial electron energy in the pre-shock flow, which is shown as a dashed blue line in the inset (a) for comparison. In the bottom panel of \fig{spectimemi}, the energy equipartition between ions and electrons  at $\ompt=2500$ explains the proximity of the thermal peaks in the electron (red solid line) and ion (red dotted line) spectra.

Since electrons and ions enter the shock with nearly the same energy, the injection efficiency into the acceleration process will be similar for the two species. In fact, by comparing the red solid line (for electrons) and the red dotted line (for ions) in the bottom panel of \fig{spectimemi}, we show that the non-thermal tails in the electron and ion spectra have comparable normalizations, yielding an acceleration efficiency for both  species of roughly $\sim1\%$ by number and $\sim10\%$ by energy. The power law  in the ion non-thermal tail is steeper than for electrons ($p\simeq3.0$ for ions and $p\simeq2.5$ for electrons), but it may evolve toward flatter slopes at later times, so that for both species the spectral index  will asymptote to the value $p\simeq2.5$ found in electron-positron shocks. 

For both ions and electrons, the non-thermal tail extends in time to higher and higher energies. As shown in the inset (b) of  \fig{spectimemi}, the maximum energy at late times increases as $\propto(\ompt)^{1/2}$, for both ions (red line) and electrons (blue line).\footnote{\citet{martins_09} reported a different scaling for the temporal evolution of the energy of shock-accelerated particles, but their conclusions might have been misguided by the limited timespan of their simulation runs.} This is the same scaling observed in \fig{spectime0} for electron-positron shocks. The similarity between electron-positron and electron-ion shocks is to be expected, since in  electron-ion flows the two species enter the shock nearly in energy equipartition, so our electron-ion shocks should display the same acceleration physics present in electron-positron flows.  From the inset (b), we find that at late times ($\ompt\gtrsim700$)
\be\label{eq:gmax3}
\frac{\gamma_{max,i}}{\gamma_0}\sim\frac{\gamma_{max,e} m_e}{\gamma_0m_i}\simeq0.25\,(\ompt)^{1/2}~~.
\ee
Even though the scaling of the maximum energy as $\propto \gamma_0 (\ompt)^{1/2}$ is the same as in electron-positron shocks (see \eq{gmax2}), the coefficient of proportionality in electron-ion shocks is smaller by a factor of $\sim2$. This  simply reflects  the position of the thermal peak, located at $\gamma_{pk}\sim\gamma_0$ in electron-positron flows and at $\gamma_{pk,i}\sim \gamma_{pk,e}m_e/m_i\sim\gamma_0/2$ in electron-ion flows, just as a result of efficient ion-to-electron energy transfer ahead of the shock (see the location of the  thermal peaks in \fig{spectimemi}, as compared to the bulk Lorentz factor $\gamma_0=15$). 

With a similar argument, we expect that in electron-ion flows with magnetization $\sigma$ the maximum Lorentz factor should saturate at
\be\label{eq:gsat3}
\frac{\gamma_{sat,i}}{\gamma_0}\sim\frac{\gamma_{sat,e} m_e}{\gamma_0m_i}\simeq\frac{2}{\sigma^{1/4}}~~,
\ee
which replaces \eq{gsat2}, valid for electron-positron plasmas. For $\sigma=\ex{5}$, the expected saturation value would be $\gamma_{sat,i}\sim \gamma_{sat,e}m_e/m_i\simeq550$, well beyond the maximum Lorentz factor reached at the end of our simulation. We would need to evolve the simulation up to much longer times ($\ompt\sim18000$) to capture the saturation of the Lorentz factor in our $\sigma=\ex{5}$ electron-ion shock.

We then extend our investigation to different magnetizations, in the range $0\lesssim\sigma\lesssim\ex{3}$. In \fig{specsigmi}, we show how the post-shock spectrum of ions (top panel) and electrons (bottom panel) depend on the magnetization of the flow. We find that shocks with $\sigma\lesssim3\times\ex{5}$ are efficient particle accelerators, and the spectrum shows a prominent non-thermal tail containing $\sim1\%$ of particles and $\sim10\%$ of flow energy (for both ions and electrons). The exponential cutoff of the non-thermal tail extends in time to higher energies, and at late times it follows the scaling $\propto(\ompt)^{1/2}$ discussed above,  for both ions (inset (a)) and electrons (inset (b)). The coefficient of proportionality tends to be lower for higher $\sigma$ (compare the green line, for $\sigma=3\times\ex{5}$, to the red line, for $\sigma=0$). This minor dependence on $\sigma$ might be caused by the slight decrease in the transverse wavelength $\lambda$ of the Weibel modes with increasing magnetization, as apparent in \fig{fluidmi} (compare panels (a), (d) and (g) for $\sigma=0$, $\sigma=\ex{5}$ and $\sigma=3\times\ex{5}$, respectively). In turn, the transverse scale $\lambda$ of the Weibel fluctuations enters the temporal scaling of the maximum energy, as illustrated in \eq{gmax1}. However, the effect of the magnetization on the scaling of the maximum energy is of minor importance, and we will neglect it in the following.

With increasing magnetization, the acceleration efficiency decreases (blue lines for $\sigma=\ex{4}$), and for $\sigma\gtrsim\ex{3}$ the post-shock spectrum does not show any evidence for non-thermal particles. For $\sigma=\ex{3}$ (black lines), the downstream electrons populate a thermal distribution, that peaks at a lower energy than for smaller magnetizations, since the Weibel modes that mediate the electron heating ahead of $\sigma\lesssim\ex{4}$ shocks are now suppressed. The non-Maxwellian shape of the ion spectrum for $\sigma=\ex{3}$ results from incomplete ion thermalization at the shock front, and it will eventually relax to a Maxwellian distribution further downstream from the shock. The absence of non-thermal particles for $\sigma=\ex{3}$ proves that shock rippling in electron-ion relativistic flows (see \fig{fluidmi}(m)) is not a promising mechanism to mediate the injection of particles into the Fermi process. In the insets (a) and (b) of \fig{specsigmi}, the suppression of non-thermal particle acceleration in $\sigma\gtrsim\ex{4}$ shocks is revealed by the fact that the maximum Lorentz factor of ions (panel (a)) and electrons (panel (b)) is nearly constant with time (blue and black lines).

Finally, we refer to SS11 for a detailed investigation of the acceleration physics of electron-ion shocks at higher magnetizations ($\sigma\gtrsim\ex{2}$), a regime where the electromagnetic precursor wave resulting from the synchrotron maser instability \citep{hoshino_91,hoshino_08} can appreciably affect the pre-shock flow.


\begin{figure}[tbp]
\begin{center}
\includegraphics[width=0.5\textwidth]{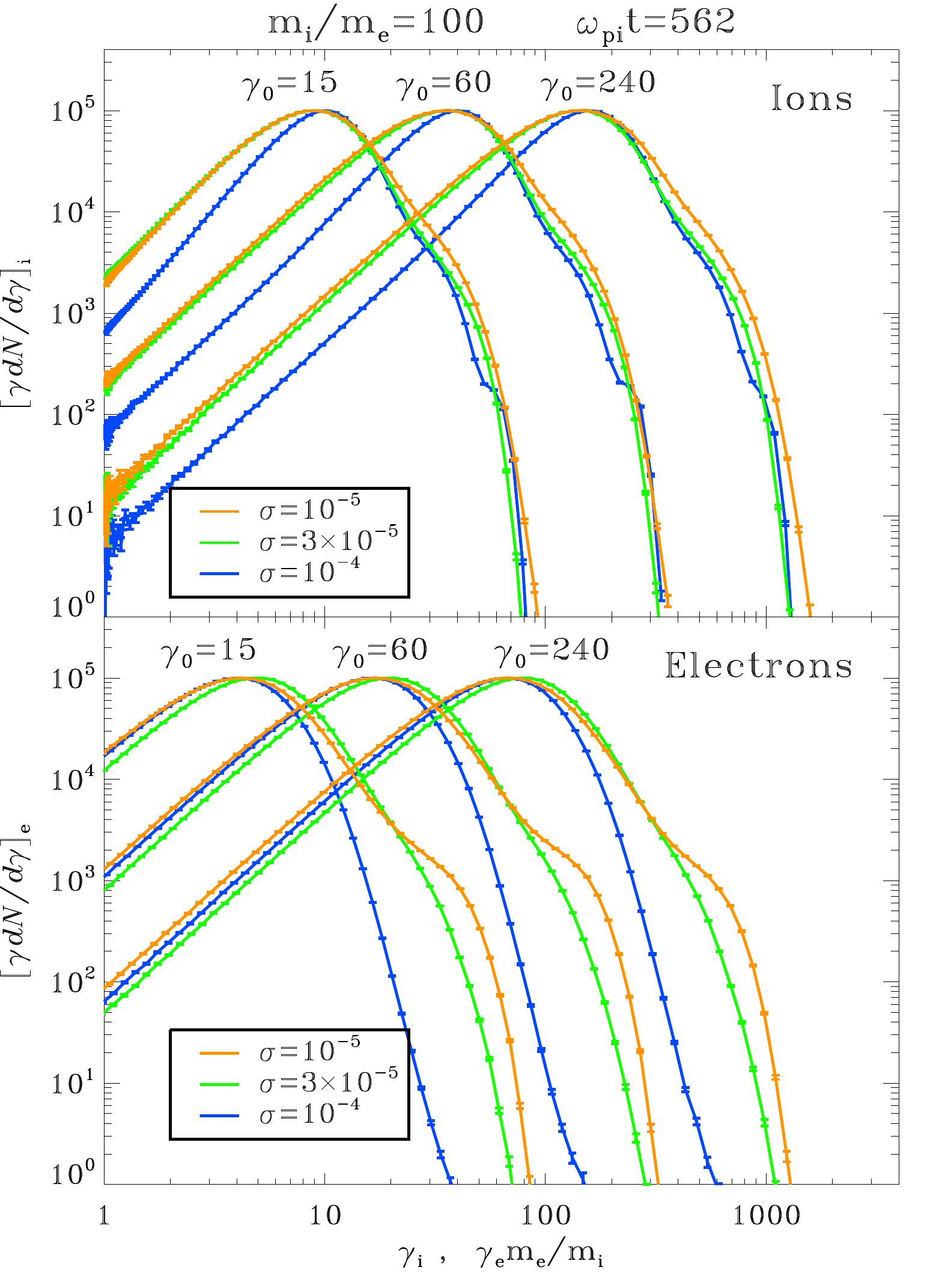}
\caption{Dependence of the post-shock particle spectrum on the upstream bulk Lorentz factor $\gamma_0$, from a set of 2D simulations of electron-ion ($\mime=100$) shocks with different magnetizations (yellow for $\sigma=\ex{5}$, green for $\sigma=3\times\ex{5}$, blue for $\sigma=\ex{4}$). In the top panel we show the ion spectrum, in the bottom panel the electron spectrum. }
\label{fig:specgammi}
\end{center}
\end{figure}

\subsection{Dependence on the Lorentz Factor $\gamma_0$}\label{sec:specgammi}
\fig{specgammi} shows how the spectrum of electron-ion ($\mime=100$) magnetized shocks depends on the bulk Lorentz factor $\gamma_0$, by comparing the cases $\gamma_0=60$ and 240 with our reference value $\gamma_0=15$.  For both ions (top panel) and electrons (bottom panel), we find that our results are basically the same regardless of $\gamma_0$, modulo an overall shift in the energy scale. \fig{specgammi} confirms that shocks with $\sigma=\ex{4}$ are poor particle accelerators (blue lines), whereas the generation of a prominent non-thermal tail is observed for $\sigma\lesssim3\times\ex{5}$ (green line for $\sigma=3\times\ex{5}$, yellow line for $\sigma=\ex{5}$). 

Most importantly, the spectra in \fig{specgammi}, obtained for a mass ratio $\mime=100$, are entirely consistent with the results in \fig{specsigmi}, referring to $\mime=25$. This suggests that the acceleration physics in electron-ion shocks can be confidently captured with a reduced mass ratio, as we will further demonstrate in Appendix \ref{sec:specmime}.


\section{Astrophysical Applications}\label{sec:apply}
In this section, we explore the implications of our results for the acceleration process at the external shocks of GRB afterglows (in \S\ref{sec:grb}) and at the termination shock of Pulsar Wind Nebulae (PWNe; in \S\ref{sec:pwn}). Our discussion of the termination shock in pulsar winds might equally apply to the termination hot spots of AGN jets -- for example, the parameters of the Cygnus A hot spots  \citep{stawarz_07} are surprisingly similar to the termination shock of the Crab Nebula.

We make use of the results presented in the previous sections. In particular, we employ the fact that the evolution in time of the Lorentz factor of shock-accelerated particles follows the scalings
\be\label{eq:gmax4a}
&\gamma_{max}&\simeq0.5\,\gamma_0\,(\ompt)^{1/2}\\
\gamma_{max,i}&\sim\frac{\gamma_{max,e} m_e}{m_i}&\simeq0.25\,\gamma_0\,(\ompt)^{1/2}\label{eq:gmax4b}
\ee
in electron-positron and electron-ion shocks, respectively. Here, $\gamma_0$ is the bulk Lorentz factor of the upstream flow (in the post-shock frame) and $\omega_{\rm pi}$ indicates either the positron plasma frequency (for electron-positron flows) or the ion plasma frequency (for electron-ion shocks).

The increase of the maximum particle energy over time proceeds up to a saturation Lorentz factor that is constrained by the magnetization $\sigma$ of the upstream flow
\be\label{eq:gsat4a}
&\gamma_{sat}&\simeq4\,\gamma_0\,\sigma^{-1/4}\\
\gamma_{sat,i}&\sim\frac{\gamma_{sat,e} m_e}{m_i}&\simeq2\,\gamma_0\,\sigma^{-1/4}\label{eq:gsat4b}
\ee
in electron-positron and electron-ion shocks, respectively. Further energization is prevented by the fact that the self-generated turbulence is confined within a region of thickness $L_{B,sat}\sim \sigma^{-1/2} \comp$ around the shock.

\subsection{GRB Afterglows}\label{sec:grb}
The afterglow emission in GRBs  is usually attributed to synchrotron radiation from electrons accelerated at the forward shock \citep[the so-called ``jitter'' paradigm does not seem to relevant in GRB afterglows, as shown by][]{sironi_spitkovsky_09b}. The shock propagates either in the ISM, with constant density, or in the stellar wind of the GRB progenitor, with density scaling in radius as $n\propto R^{-2}$. As discussed in \S\ref{sec:intro}, the magnetization of the ISM is $\sigma\sim\ex{9}$, and it is independent of the shock radius $R$. The same holds for a stellar wind, if the magnetic field in the wind is primarily toroidal, thus decreasing as $\propto R^{-1}$. We assume that, as appropriate for a Wolf-Rayet progenitor \citep{eichler_usov_93,sagi_nakar_12}, the number density in the wind at $R=10^{18}\unit{cm}$ is $n\equiv 0.3\, n_{-0.5} \unit{cm^{-3}}$, and the strength of the magnetic field at $R=10^{18}\unit{cm}$  is $B_{\rm W}\equiv \ex{5} B_{\rm W,-5}\unit{G}$, so that the magnetization parameter in the wind will be
\be
\sigma=\frac{B_{\rm W}^2}{4\pi n m_i c^2}\simeq1.7\times\ex{8}B_{\rm W,-5}^2 n_{-0.5}^{-1}~.
\ee
We treat separately the case of a blast wave propagating into the ISM  (\S\ref{sec:ism}) and into a stellar wind (\S\ref{sec:wind}).

In the following, we study the process of particle acceleration in GRB afterglows by assuming that the shock properties are only determined by the instantaneous magnetization and Lorentz factor of the flow. In other words, we neglect the feedback of  particles and magnetic turbulence produced at early times on the shock structure and the acceleration process at later times. 

Also, we only focus on the particles reaching the highest energies, with the goal of assessing whether GRB afterglow shocks are promising candidates for the acceleration of the UHECR protons observed by the Pierre Auger Observatory \citep[][]{auger_10}, and whether synchrotron radiation from the shock-accelerated electrons can explain the early sub-GeV emission of  Fermi-LAT GRBs \citep[e.g.,][]{ackermann_10,depasquale_10,ghisellini_10}. Our strategy parallels closely the study by \citet{sagi_nakar_12}, with one important difference. They were inferring the rate of electron acceleration in GRB external shocks from the detection of high-energy photons at early times, whereas our PIC simulations can provide from first principles an estimate for the acceleration rate. Our goal is then to verify whether  synchrotron emission from the accelerated electrons can explain the highest energy photons detected by Fermi-LAT.

Since we only focus on the highest energy electrons, we do not attempt to provide a complete description of the full spectral and temporal evolution of GRB afterglows. For this, we would need a self-consistent estimate for the overall extent of the region populated with Weibel turbulence. A strict lower limit is provided by the thickness $L_{B,sat}$ of the \tit{upstream} layer filled with Weibel filaments (see \eq{foot}), but the decay length of the Weibel-generated fields in the \tit{downstream} region is still a matter of debate \citep{chang_08,keshet_09}. A physically-grounded estimate for the overall extent of the region with Weibel turbulence could  be used to predict the temporal and spectral evolution of GRB afterglows, following \citet{rossi_rees_03} and \citet{lemoine_12}.

\subsubsection{ISM}\label{sec:ism}
Under the assumption of adiabatic expansion in a constant density medium,  the bulk Lorentz factor of the shock during the relativistic deceleration phase will be described by the \citet{blandford_76} solution
\be
\Gamma(R)=\left(\frac{17 E_0}{16\pi n m_i c^2}\right)^{1/2} R^{-3/2}~,
\ee
at distances larger than the deceleration radius $R_{dec}\equiv(17 E_0/16\pi \Gamma_0^2 n m_i c^2)^{1/3}$. Here, $E_0$ is the isotropic-equivalent explosion energy, $\Gamma_0$ is the initial Lorentz factor of the blast wave, and $n$ is the ISM number density  (measured in the rest frame of the ISM). By defining $E_0\equiv10^{54}E_{0,54}\unit{ergs}$, $\Gamma_0\equiv 10^{2.5}\Gamma_{0,2.5}$ and $n\equiv n_{0}\unit{cm^{-3}}$, the deceleration radius amounts to 
\be
R_{dec}\simeq 1.3\times 10^{17} E_{0,54}^{1/3} \Gamma_{0,2.5}^{-2/3} n_0^{-1/3}\unit{cm}~.
\ee 
The shock becomes non-relativistic at $R_{nr}=(17 E_0/16\pi n m_i c^2)^{1/3}\simeq6.1\times 10^{18} E_{0,54}^{1/3}n_0^{-1/3}\unit{cm}$, at which point our results will no longer be applicable.

We now evaluate the acceleration performance of GRB external shocks, first for protons and then for  electrons. We use  the fact that the Lorentz factor $\gamma_0$ we have employed in the previous sections coincides with the instantaneous Lorentz factor of the GRB blast wave, i.e., $\gamma_0=\Gamma$ at all radii.  The maximum Lorentz factor of the shock-accelerated protons  is constrained by the magnetization of the pre-shock flow, as illustrated in \eq{gsat4b}. When transforming into the upstream frame (coincindent with the ISM frame), the result in \eq{gsat4b} needs to be multiplied by an additional factor of  $\Gamma$, giving
\be\label{eq:prot1}
\gamma^{\rm up}_{sat,i}\simeq 8.0\times10^7\,E_{0,54}\,n_0^{-1}\,\sigma_{-9}^{-1/4}R_{17}^{-3}~,
\ee
where $\sigma\equiv\ex{9}\sigma_{-9}$ and $R\equiv10^{17}R_{17}\unit{cm}$. This strict upper limit could be circumvented if the coherence scale $\lambda_{coh}$ of the ISM magnetic field in the longitudinal direction (i.e., along the shock normal) is small enough such that the acceleration of protons up to $\gamma^{\rm up}_{sat,i}$ takes longer than the passage of the shock through a region of length $\lambda_{coh}$. From the acceleration timescale in \eq{gmax4b}, this argument gives an upper limit on the required longitudinal coherence length of the ISM field (in the ISM frame)
\be\label{eq:lcoh1}
\lambda_{coh}\lesssim 2.2\times10^{16} E_{0,54}^{1/2} n_{0}^{-1}\sigma_{-9}^{-1/2}R_{17}^{-3/2}\unit{cm}~,
\ee
where in \eq{gmax4b} we have used the fact that the proton plasma frequency in the ISM is
\be
\omega_{\rm pi}\simeq 1.3\times10^3\, n_0^{1/2}\unit{Hz}~.
\ee
The coherence scale of the ISM magnetic field is believed to be a few parsecs, so the constraint in \eq{lcoh1} is not likely to be satisfied, and proton acceleration to Lorentz factors larger than $\gamma^{\rm up}_{sat,i}$ will not occur in the ISM.

A separate constraint comes from the requirement that the proton acceleration timescale in \eq{gmax4b} should be shorter than the age of the blast wave $\sim R/\Gamma c$ as measured in the post-shock frame. 
This constrains the maximum proton Lorentz factor to be, in the ISM  frame,
\be\label{eq:prot2}
\gamma^{\rm up}_{age,i}\simeq1.7\times10^{8}E_{0,54}^{3/4} n_0^{-1/2}R_{17}^{-7/4}\!\!\!~.
\ee
By comparing \eq{prot1} and \eq{prot2}, we see that the maximum proton Lorentz factor $\gamma^{\rm up}_{max,i}\equiv\min[\gamma^{\rm up}_{sat,i},\gamma^{\rm up}_{age,i}]$ is always limited by the magnetization of the pre-shock flow, as prescribed in \eq{prot1}. From this, it is clear that the maximum energy of protons accelerated in GRB external shocks propagating into the ISM is too small to explain the extreme Lorentz factors ($\sim10^{11}$) of the UHECRs  observed by Auger \citep[][]{auger_10}.

We now turn to the acceleration of electrons. The Lorentz factor of the thermal peak in the electron distribution will be, in the post-shock frame,
\be\label{eq:peak}
\gamma_{pk,e}\sim\frac{\Gamma m_i}{2 m_e}\simeq4.4\times 10^5\,E_{0,54}^{1/2} n_0^{-1/2}R_{17}^{-3/2}~,
\ee
where the factor of two in the denominator comes from the assumption of energy equipartition between electrons and protons behind the shock. The Lorentz factor $\gamma_{pk,e}$ is a strict lower limit for the low-energy end of the non-thermal tail (or equivalently,  for the minimum Lorentz factor $\gamma_{min,e}$ of the shock-accelerated electrons), so we suggest to take $\gamma_{min,e}\sim\gamma_{pk,e}$.

With regards to the acceleration of electrons,  \eq{gsat4b}  gives an absolute upper limit constrained by the flow magnetization, which in the post-shock frame yields
\be\label{eq:lecs1}
\gamma_{sat,e}\simeq 3.1\times10^8\,E_{0,54}^{1/2} n_0^{-1/2}\sigma_{-9}^{-1/4}R_{17}^{-3/2}~.
\ee
As explained above, this strict upper limit implicitly assumes that the ISM field does not change its orientation as electrons are accelerated up to the Lorentz factor $\gamma_{sat,e}$, which is generally expected given the parsec-scale coherence length of the ISM magnetic field.

Radiative energy losses are likely to limit the maximum electron Lorentz factor to  values smaller than $\gamma_{sat,e}$. The electrons reaching the most extreme energies will have to remain in the acceleration region throughout their life, so that they will preferentially cool in the Weibel-generated fields, rather than in the background field $B_0$ (at odds with the assumption by \citet{kumar_12}). By comparing \eq{gmax1} and \eq{gmax4b}, we can constrain the combination $\lambda_{\comp}\epsilon_B\simeq0.03$. By assuming that the characteristic transverse wavelength of the Weibel fluctuations is $\lambda\simeq10\comp$ (or equivalently, $\lambda_{\comp}\simeq10$), we can estimate $\epsilon_B\simeq3\times\ex{3}$ as the mean value of the magnetic energy fraction in the acceleration region.\footnote{From now on, we will keep $\epsilon_B$ as a free parameter. However, the combination $\lambda_{\comp}\epsilon_B\simeq0.03$ needs to stay fixed, so we expect $\epsilon_B$ to vary by at most a factor of ten, as a result of different choices for the transverse scale $3\lesssim\lambda_{\comp}\lesssim30$ of the Weibel filaments.} This is much larger than $\sigma$, suggesting that the electrons will primarily cool due to the self-generated fields (we will check this \tit{a posteriori}). By balancing the acceleration time in \eq{gmax4b} with the synchrotron cooling time in the Weibel-generated fields, we infer the cooling-limited Lorentz factor in the post-shock frame\footnote{Here, we only consider synchrotron cooling. To include IC losses, one would need to generalize the arguments in \citet{li_waxman_06,piran_nakar_10,li_zhao_11,sagi_nakar_12} using the scalings we have found in \eq{gmax4b} and \eq{gsat4b}.}
\be\label{eq:lecs2}
\gamma_{sync,e}\simeq1.2\times10^7\,n_0^{-1/6} \epsilon_{B,-2.5}^{-1/3}
\ee
where $\epsilon_B\equiv\ex{2.5} \epsilon_{B,-2.5}$. Remarkably, the limit in \eq{lecs2} does not depend on the shock radius. By comparing \eq{lecs1} with \eq{lecs2}, we see that synchrotron losses limit the maximum electron Lorentz factor $\gamma_{max,e}\equiv\min[\gamma_{sat,e},\gamma_{sync,e}]$ only at small radii, whereas for $R\gtrsim\,8.5\times 10^{17}\unit{cm}$ the effect of the flow magnetization in \eq{lecs1} becomes more constraining. 

For electrons, the requirement that the acceleration time in \eq{gmax4b} should be shorter than the age of the blast wave ($\sim R/\Gamma c$) yields a critical Lorentz factor
\be
\gamma_{age,e}\simeq6.6\times10^{8}E_{0,54}^{1/4}R_{17}^{-1/4}~,
\ee
which is always larger (so, less constraining) than $\gamma_{max,e}=\min[\gamma_{sat,e},\gamma_{sync,e}]$ defined above.

A posteriori, we can check the validity of our initial assumption, that the highest energy electrons primarily cool due to the Weibel-generated fields, rather than the background field $B_0$. As discussed by \citet{kumar_12}, this is realized if
\be\label{eq:kumar}
\epsilon_B\, L_{B,sat} \gg \sigma\, r_{L,\sigma}(\gamma_{max,e})
\ee
where $r_{L,\sigma}(\gamma_{max,e})=(\gamma_{max,e}m_e/\gamma_0 m_i) \,\sigma^{-1/2}\comp$ is the Larmor radius in the background field $B_0$, and we have taken $L_{B,sat}$ from \eq{foot} as a conservative lower limit for the thickness of the region filled with Weibel turbulence around the shock (in reality, $L_{B,sat}$ only accounts for the extent of the turbulent region on the upstream side of the shock). In \eq{kumar}, the left-hand side quantifies the energy lost to synchrotron emission as the accelerated particles propagate across a region of extent $\sim L_{B,sat}$ filled with Weibel fields of energy density $\propto \epsilon_B$. In contrast, the right-hand side estimates the cooling losses due to the background field (with energy density $\propto \sigma$) during one Larmor period of the highest energy electrons. From the expression for $L_{B,sat}$ in \eq{foot}, we can rewrite the previous inequality as 
\be\label{eq:cool}
\epsilon_B/\sigma\gg \gamma_{max,e}/\gamma_{pk,e}
\ee
where we have assumed $\eta_{inj}=1$ in \eq{foot} for simplicity, and we have used that $\gamma_{pk,e}\sim \gamma_0 m_i/m_e$. For our typical parameters ($\epsilon_B\simeq3\times\ex{3}$, $\sigma=\ex{9}$, $\gamma_{pk,e}$ in \eq{peak}, and $\gamma_{max,e}$ from the minimum between \eq{lecs1} and \eq{lecs2}), the inequality in \eq{cool} is easily satisfied, for all Lorentz factors $\leq\gamma_{max,e}$.\footnote{\citet{kumar_12} came to the opposite conclusion, because they (incorrectly) employed a value of $L_{B,sat}\sim \comp$ that is almost five orders of magnitude smaller than what we find based on the results of our PIC simulations.}

It is also straightforward to verify that $\epsilon_B L_{B,sat}\gtrsim \sigma R/\Gamma$ during the whole relativistic deceleration phase, a criterion that is even more constraining than \eq{kumar}. Here, we are comparing the cooling losses in the Weibel fields (across a region of thickness $\sim L_{B,sat}$) with the energy lost to synchrotron radiation in the background field, assuming that it permeates the whole GRB shell (of thickness $\sim R/\Gamma$, as measured in the post-shock frame). This confirms that most of the synchrotron cooling occurs in the self-generated Weibel fields.

From the expressions for $\gamma_{sync,e}$ and $\gamma_{sat,e}$ derived above, we can compute the corresponding synchrotron frequencies as a function of the observer's time. We use that the observer's time is related to the shock radius by $t_{obs}=(1+z) R/2 \Gamma^2$ \citep[e.g.,][]{waxman_97}, where $z$ is the redshift. By boosting the emitted comoving frequencies by a factor of $\sim2\,\Gamma/(1+z)$ to obtain the observed frequencies \citep{piran_nakar_10}, we obtain 
\be
\!\!\!\!\!\!\!\!\!\!\!\!\!h \nu_{sync,e}&\simeq&1.8\,E_{0,54}^{1/4}n_0^{-1/12}\!\!\epsilon_{B,-2.5}^{-1/6}(1+z)^{-1/4} t_{obs,2}^{-3/4}\unit{GeV}\label{eq:nu1}\\
\!\!\!\!\!\!\!\!\!\!\!\!\!h \nu_{sat,e}&\simeq&0.16\,E_{0,54}^{1/2}\epsilon_{B,-2.5}^{1/2}\sigma_{-9}^{-1/2}(1+z)^{1/2} t_{obs,2}^{-3/2}\unit{TeV}
\ee
where $t_{obs}\equiv10^2 t_{obs,2}\unit{s}$. The maximum energy of synchrotron afterglow photons will satisfy $h\nu_{max,e}=h \nu_{sync,e}$ at early times ($t_{obs}\lesssim4\times10^4\unit{s}$), whereas $h\nu_{max,e}=h \nu_{sat,e}$ at later times. Our results show that shock-accelerated electrons cooling in the Weibel  fields can radiate synchrotron photons up to energies of a few GeV in the early phases of GRB afterglows. This explains both the GeV photons seen $\sim100\unit{s}$ after the explosion in a  number  of Fermi-LAT GRBs \citep[e.g.,][]{ackermann_10,depasquale_10,ghisellini_10}, as well as the $>100\unit{MeV}$ photons seen in large numbers up to $\sim1000\unit{s}$ after the explosion.
  
  
\subsubsection{Stellar Wind}\label{sec:wind}
For the sake of completeness, we also consider the case of an adiabatic GRB blast wave propagating in the  wind of its progenitor star. Let us define $A=m_i n R^2$, where $n\propto R^{-2}$ is the particle density in the wind. The Lorentz factor of the blast wave will evolve as
\be
\Gamma(R)=\left(\frac{9 E_0}{16 \pi A  c^2}\right)^{1/2} R^{-1/2}
\ee
from the deceleration radius $R_{dec}\simeq6.3\times10^{15}E_{0,54} \Gamma_{0,2.5}^{-2} A_{11.5}^{-1}\unit{cm}$ up to the non-relativistic radius $R_{nr}\simeq6.3\times10^{20}E_{0,54} A_{11.5}^{-1}\unit{cm}$. Here, we have defined $A=3 \times 10^{11}A_{11.5}\unit{g\,cm^{-1}}$, as appropriate for the dense wind of Wolf-Rayet stars.

For protons,  \eq{gsat4b} states that the magnetization of the stellar wind constrains the maximum Lorentz factor in the wind frame to be
\be\label{eq:gsatw}
\gamma^{\rm up}_{sat,i}\simeq 1.3\times10^7\,E_{0,54}\,A_{11.5}^{-1}\,\sigma_{-8}^{-1/4}R_{16}^{-1}~,
\ee
where $\sigma\equiv10^{-8}\sigma_{-8}$ and $R\equiv 10^{16}R_{16}\unit{cm}$. Using the same argument as in \S\ref{sec:ism}, acceleration to higher energies is possible if the longitudinal coherence length of the wind magnetic field satisfies (in the wind frame)
\be\label{eq:lcohw}
\lambda_{coh}\lesssim 8.4\times10^{13} E_{0,54}^{1/2} A_{11.5}^{-1}\sigma_{-8}^{-1/2}R_{16}^{1/2}\unit{cm}~,
\ee
so that the shock can cross a region of length $\lambda_{coh}$ before the protons are accelerated up to $\gamma^{\rm up}_{sat,i}$. For the acceleration timescale in \eq{gmax4b}, we have used that the proton plasma frequency in the wind is
\be
\omega_{\rm pi}\simeq 5.7\times10^4\, A_{11.5}^{1/2}R_{16}^{-1}\unit{Hz}~.
\ee
In the equatorial plane of the stellar wind, the polarity of the toroidal magnetic field will alternate with the stellar rotation period. The half-wavelength of the Parker spiral will be $\lambda_{\rm W}\sim \pi (v_{\rm W}/v_{rot}) R_\star$, and for typical parameters of Wolf-Rayet stars \citep[ratio of wind velocity to stellar rotation velocity $v_{\rm W}/v_{rot}\sim10$, and stellar radius $R_\star\sim20 R_\odot$, see][]{eichler_usov_93}, we find $\lambda_{W}\simeq4.4\times 10^{13}\unit{cm}$, which satisfies the condition in \eq{lcohw}. It follows that acceleration of protons to energies larger than in \eq{gsatw} is generally allowed in the equatorial plane of the stellar wind.

A separate constraint comes from the age of the blast wave ($\sim R/\Gamma c$), which limits the maximum proton Lorentz factor to be, in the wind frame,
\be
\gamma^{\rm up}_{age,i}\simeq1.4\times10^{8}E_{0,54}^{3/4} A_{11.5}^{-1/2}R_{16}^{-3/4}\!\!\!~,
\ee
which is a strict upper limit that does not depend on the coherence scale of the circum-burst magnetic field.

With regards to electrons, the thermal peak of their energy spectrum, which we take as a proxy for the low-energy end of the electron power-law tail, will be
\be
\gamma_{pk,e}\sim\frac{\Gamma m_i}{2 m_e}\simeq2.3\times 10^5\,E_{0,54}^{1/2} A_{11.5}^{-1/2}R_{16}^{-1/2}~.
\ee
From \eq{gsat4b}, the magnetization of the stellar wind constrains the maximum electron Lorentz factor to be
\be\label{eq:wind3}
\gamma_{sat,e}\simeq 9.2\times10^7\,E_{0,54}^{1/2} A_{11.5}^{-1/2}\sigma_{-8}^{-1/4}R_{16}^{-1/2}~.
\ee
Another constraint comes by balancing the acceleration time in \eq{gmax4b} with the synchrotron cooling time in the Weibel fields, which gives
\be\label{eq:wind4}
\gamma_{sync,e}\simeq3.5\times10^6\,A_{11.5}^{-1/6} \epsilon_{B,-2.5}^{-1/3} R_{16}^{1/3}~,
\ee 
so that the maximum electron Lorentz factor is set initially by $\gamma_{max,e}=\gamma_{sync,e}$, it increases in radius as $\propto R^{1/3}$ (see \eq{wind4}) up to $R\simeq5.0\times10^{17}\unit{cm}$, and for larger distances it decreases as $\propto R^{-1/2}$ following $\gamma_{sat,e}$ in \eq{wind3} (but it may be even larger than $\gamma_{sat,e}$ in the equatorial plane of the wind, where the field orientation changes with the stellar rotation period). As observed in \S\ref{sec:ism} for the ISM case, the age of the blast wave does not play a major role in limiting the maximum electron energy. From the expression for $\gamma_{max,e}=\min[\gamma_{sat,e},\gamma_{sync,e}]$ obtained above, it is easy to verify that $\epsilon_B/\sigma\gg \gamma_{max,e}/\gamma_{pk,e}$ even in the case of a wind-like medium (see \eq{cool} for the ISM profile), so  the highest energy electrons primarily cool in the Weibel-generated fields, rather than in the background field.

From the expressions for $\gamma_{sync,e}$ and $\gamma_{sat,e}$, we obtain the maximum  energies of synchrotron photons
\be
\!\!\!\!\!\!\!\!\!\!\!\!\!h \nu_{sync,e}&\simeq&1.1\,E_{0,54}^{1/3}A_{11.5}^{-1/6}\epsilon_{B,-2.5}^{-1/6}(1+z)^{-1/3} t_{obs,2}^{-2/3}\unit{GeV}\\
\!\!\!\!\!\!\!\!\!\!\!\!\!h \nu_{sat,e}&\simeq&37.1\,E_{0,54}^{1/2}\epsilon_{B,-2.5}^{1/2}\sigma_{-8}^{-1/2}(1+z)^{1/2} t_{obs,2}^{-3/2}\unit{GeV}
\ee
The maximum energy of  afterglow photons will satisfy $h\nu_{max,e}=h \nu_{sync,e}$ at early times ($t_{obs}\lesssim7\times10^3\unit{s}$), whereas $h\nu_{max,e}=h \nu_{sat,e}$ at later times. Synchrotron emission from the external shock of a GRB blast wave propagating in a  stellar wind is then a valid candidate for producing the sub-GeV signature of Fermi-LAT bursts.

The expression derived above for $h \nu_{sat,e}$ can be used to interpret the very sharp temporal break observed at $t_{obs}\sim4\times 10^7\unit{s}$ in the light curve of GRB 060729, the burst with the latest Chandra X-ray detection. \citet{grupe_10} found that the model of a spherical blast wave with $E_{0,54}=1$ propagating in a stellar wind with $A_{11.5}=0.1$ fits the data very well until the temporal break  at $t_{obs}\sim4\times 10^7\unit{s}$. Following the suggestion by \citet{sagi_nakar_12}, we claim that the temporal break at late times (which has been tentatively associated to a softening in the 0.3-10 keV spectrum) could be caused by the inability of the shock to accelerate X-ray emitting particles, due to the limitation given by the wind magnetization. At $t_{obs}\sim4\times 10^7\unit{s}$, the shock is still relativistic ($\Gamma\simeq7$), so our results are still applicable. By imposing $h \nu_{sat,e}\simeq0.3\unit{keV}$ at $t_{obs}\sim4\times 10^7\unit{s}$ for a burst at $z=0.54$, as appropriate for GRB 060729, we constrain the magnetization of the stellar wind to be $\sigma\simeq3\times10^{-9}$.

\subsection{Pulsar Wind Nebulae}\label{sec:pwn}
The spectrum of Pulsar Wind Nebulae consists of two components, where the low energy component, most likely dominated by synchrotron, shows a cutoff at a few tens of MeV. The fact that synchrotron emission reaches these energies, despite the rapid synchrotron cooling, implies that particle acceleration in the nebula is an extremely fast process \citep{dejager_harding_92}. In this section, we study the acceleration properties of the termination shock in pulsar winds, taking the Crab Nebula as our prototypical example.

Around the equatorial plane of obliquely-rotating pulsars, the wind consists of toroidal stripes of opposite magnetic polarity, separated by current sheets of hot plasma. It is still a subject of active research whether the alternating stripes will dissipate their energy into particle heat ahead of the termination shock, or whether the wind remains dominated by Poynting flux till the termination shock \citep[][]{lyubarsky_kirk_01,kirk_sk_03}. \citet{lyubarsky_03} showed that, as regards to the post-shock properties of the flow, the effect of the stripe dissipation at the termination shock is equivalent to their annihilation well ahead of the shock. However, when considering the synchrotron losses of the accelerated particles in the pre-shock fields, it does make a difference whether or not the stripes are surviving till the termination shock. Here, we  only consider the  case where the stripes are dissipated far ahead of the termination shock.

The pulsar wind is an electron-positron flow, moving towards the termination shock with a relativistic bulk Lorentz factor \citep[see][]{lyubarsky_03}
\be
\gamma_0\sim \frac{L_{sd}}{m_e c^2 \dot{N}}\simeq3.7\times 10^{4} L_{sd,38.5}\dot{N}_{40}^{-1}~,
\ee
where we have assumed complete dissipation of the alternating stripes in the wind. Here, $L_{sd}\equiv 3 \times 10^{38}L_{sd,38.5}\unit{erg s\,s^{-1}}$ is the spin-down luminosity of the Crab, and $\dot{N}=10^{40}\dot{N}_{40}\unit{s^{-1}}$ is the total particle flux entering the nebula, including the long-lived radio-emitting electrons \citep{bucciantini_11}. More precisely, a particle flux of $10^{40}\unit{s^{-1}}$ is a strict lower limit for radio-emitting electrons, since this estimate only accounts for the pairs needed to produce the Crab radio  emission at frequencies above the ionospheric cutoff -- obviously the radio spectrum must extend to lower frequencies, therefore the particle distribution goes to lower energies, and the inferred number of electrons will increase as the low-energy cutoff recedes. On the other hand, one could speculate that radio-emitting electrons (whose flux is $\dot{N}\gtrsim10^{40}\unit{s^{-1}}$) are only injected in the polar flow, and that the particle flux in the equatorial region, where  the  electrons emitting in the optical, X-ray and gamma-ray bands are accelerated, will be lower ($\dot{N}\sim3\times10^{38}\unit{s^{-1}}$), as argued by \citet{arons_98}. With these uncertainties in mind, we take $\dot{N}=10^{40}\unit{s^{-1}}$ as our reference value.

Even in the case of complete dissipation of the alternating fields, the pulsar wind away from the midplane will carry a net ordered field, resulting from the fact that above and below the equatorial plane the stripes are not symmetric (i.e., the widths of the regions with opposite fields are not equal). If $\zeta\simeq45^\circ$ is the inclination angle between the rotational and magnetic axis of the Crab pulsar, inferred from fitting the spectrum and pulse profile of the high-energy pulsar emission \citep{harding_08}, and if $\lambda_\zeta$ is the latitude angle measured from the midplane, the residual wind magnetization near the equatorial plane will be \citep{komissarov_12}
\be
\sigma(\lambda_\zeta)\sim\left(\frac{\tan\lambda_\zeta}{\tan\zeta}\right)^2\equiv \alpha^2~,
\ee
where we have assumed $\alpha\ll1$. Based on the results of \S\ref{sec:spec}, we require $\sigma\lesssim10^{-2}$ for efficient particle acceleration in electron-positron flows, which corresponds to $\alpha\lesssim0.1$. This is the same constraint found by \citet{sironi_spitkovsky_11b} with PIC simulations of striped pulsar winds, even though they assumed that the magnetic stripes survive till the termination shock. In other words, we find that the limit on $\alpha$ for efficient Fermi acceleration does not depend on the location where the magnetic stripes dissipate (either in the wind, or at the shock).

As in the case of GRB afterglows, an important constraint for the maximum energy of shock-accelerated particles comes from the flow magnetization. By defining $\alpha\equiv10^{-4}\alpha_{-4}$, eq.~(\ref{eq:gsat4a}) gives
\be
\gamma_{sat,e}(\alpha)\simeq1.5\times 10^7L_{sd,38.5}\dot{N}_{40}^{-1}\alpha_{-4}^{-1/2}~,
\ee
which does not provide any constraint in the  midplane (where $\alpha=0$). Here, we have adopted a local description of the acceleration process, such that the accelerated particles only experience the conditions at one single latitude. For high-energy electrons, this assumption is likely to break, and the acceleration process will be modified by the latitudinal dependence of the wind magnetization. Such a detailed analysis is beyond the scope of this work.  

Cooling in the self-generated Weibel fields provides an additional constraint, independent from the latitude. By balancing the acceleration time in \eq{gmax4a} with the synchrotron cooling time in the Weibel fields, we obtain 
\be
\!\!\gamma_{sync,e}\simeq3.5\times10^{8}L_{sd,38.5}^{1/6}\dot{N}_{40}^{-1/3} \epsilon_{B,-2.5}^{-1/3}R_{\rm TS,17.5}^{1/3}~.
\ee
To compute the positron plasma frequency in \eq{gmax4a}, we have used that the number density ahead of the termination shock is $n_{{\rm TS}}=\dot{N}/(4 \pi R_{\rm TS}^2 c)$, assuming an isotropic particle flux, so that the positron plasma frequency is
\be
\omega_{\rm pi}\equiv\sqrt{\frac{2 \pi n_{\rm TS} e^2}{\gamma_0 m_e}}\simeq0.11 \,L_{sd,38.5}^{-1/2}\dot{N}_{40} R_{\rm TS,17.5}^{-1}\unit{Hz}
\ee
where the termination shock radius has been written as $R_{\rm TS}\equiv3\times10^{17}R_{\rm TS,17.5}\unit{cm}$.

A third constraint, still independent from the latitude, comes from the requirement that the diffusion length of the highest energy electrons be smaller than the termination shock radius (i.e., a confinement constraint). Alternatively, the acceleration time in \eq{gmax4a} should be shorter than $R_{\rm TS}/c$, which yields the critical limit
\be
\gamma_{\mathit{conf,e}}\simeq1.9\times10^{7}L_{sd,38.5}^{3/4}\dot{N}_{40}^{-1/2}~,
\ee
which is generally more constraining than the cooling-limited Lorentz factor $\gamma_{sync,e}$ discussed above. As compared to the latitude-dependent $\gamma_{sat,e}$, we see that the maximum electron Lorentz factor will be controlled by $\gamma_{\mathit{conf,e}}$ for $\alpha_{-4}\lesssim0.6$, and by $\gamma_{sat,e}$ for higher latitudes. 

The synchrotron photons emitted by the electrons with  Lorentz factors  $\gamma_{sat,e}$ and $\gamma_{\mathit{conf,e}}$ will have energies
\be
\!\!\!h \nu_{sat,e}&\simeq&96.5\,L_{sd,38.5}^{5/2}\dot{N}_{40}^{-2}\epsilon_{B,-2.5}^{1/2}R_{\rm TS,17.5}^{-1}\alpha_{-4}^{-1}\unit{eV}\\
\!\!\!h \nu_{\mathit{conf,e}}&\simeq&0.17\,L_{sd,38.5}^{2}\dot{N}_{40}^{-1}\epsilon_{B,-2.5}^{1/2}R_{\rm TS,17.5}^{-1}\unit{keV}
\ee
 which are apparently too small to explain the X-ray spectrum of the Crab, extending to energies beyond a few MeV. This conclusion holds even in the extreme case $\epsilon_B\sim1$, which might be appropriate in computing $h \nu_{sat,e}$ and $h \nu_{\mathit{conf,e}}$ if the stripes survive till the termination shock. Moreover, even if we neglect the constraint given by $h \nu_{\mathit{conf,e}}$, we find that X-ray photons with $h \nu\gtrsim1\unit{keV}$ can  be produced only within an equatorial wedge of angular extent $\lambda_{\zeta}\lesssim\ex{5}$ (assuming an inclination $\zeta\simeq45^\circ$). The energy flux in this sector is a tiny fraction $\sim 2 \alpha\simeq2\times\ex{5}$ of the total energy flow in the wind (assuming the energy flow follows the Poynting flux of a split monopole $\propto \cos^2 \lambda_\zeta$), which is insufficient to power the Crab X-ray emission, that consumes  $\sim10\%$ of the pulsar spin-down power \citep{atoyan_96}. 
 We conclude that Fermi acceleration at the  termination shock of PWNe is not a likely candidate for producing X-ray photons via the synchrotron process, and valid alternatives should be investigated.
   
One possibility -- magnetic dissipation of the striped pulsar wind in and around the shock front itself -- has been extensively studied, with the conclusion that particle acceleration along extended X-lines formed by tearing of the current sheets may contribute to the flat particle distribution (with spectral index $p\simeq1.5$) required to explain the far infrared and radio  spectra of PWNe \citep[e.g.,][]{lyubarsky_03, sironi_spitkovsky_11b}. However, further acceleration to gamma-ray emitting energies by the Fermi process cannot occur in the transverse shock that terminates the pulsar wind, if particle scattering depends only on the self-generated turbulence. 

Yet, the steady-state  hard X-ray and gamma-ray spectra of PWNe do look like the consequences of Fermi acceleration -- particle distributions with $p \simeq 2.4$ are implied by the observations. In this regard, we argue that the wind termination shock might form in a macroscopically turbulent medium, with the outer scale of the turbulence driven by the large-scale shear flows in the nebula \citep{komissarov_04,delzanna_04,camus_09}. If the large-scale motions observed in MHD simulations of PWNe drive a turbulent cascade to shorter wavelengths, back-scattering of the particles in this downstream turbulence, along with upstream reflection by the transverse magnetic field of the wind, might sustain Fermi acceleration to higher energies. Yet, the turbulent cascade has to work in such a way as not to disturb the polarization of the nebula, which looks rather cleanly toroidal \citep{wilson_72,schmidt_79}. 

Another ``external'' influence of reconnection on the shock structure, that might lead to particle acceleration to higher energies, may be connected to the accelerator behind the recently discovered gamma-ray flares in the Crab Nebula \citep{abdo_11}. If the stripes decay well ahead of the shock, the wind has a ``Mexican hat'' magnetic geometry, with oppositely wound toroidal magnetic field in the northern and southern hemispheres, separated by the equatorial current sheet. Tearing of that current sheet can create radial spokes of current, with radially extended X-lines. Runaway acceleration of electrons and positrons at those X-lines, a linear accelerator, injects energetic beams into the shock, with the mean energy per particle approaching the whole open field line voltage, $\gtrsim 10^{16}\unit{V}$ in the Crab \citep{arons_12}.  This high-energy population can drive cyclotron turbulence when gyrating in the shock-compressed fields, and resonant absorption of the cyclotron harmonics can accelerate the electron-positron pairs in a broad spectrum, with maximum energy again comparable to the whole open field line voltage \citep{amato_arons_06}.

\section{Summary and Discussion}\label{sec:disc}
In this work, we have investigated by means of 2D and 3D PIC simulations the physics and acceleration properties of relativistic perpendicular shocks, that propagate in electron-positron or electron-ion plasmas with moderate magnetizations ($0\lesssim\sigma\lesssim\ex{1}$). As a function of the magnetization $\sigma$ and the pre-shock Lorentz factor $\gamma_0$, we have explored the efficiency  of the Fermi process (i.e., the amount of particles and energy stored in the non-thermal tail) and the rate of particle acceleration. 

We find that the Fermi process is suppressed in strongly magnetized perpendicular shocks, in agreement with the conclusions by SS09 and SS11 and with earlier 1D simulations performed by \citet{langdon_88} and \citet{gallant_92}. Due to the lack of sufficient self-generated turbulence, the charged particles are constrained to move along the field lines, which are advected downstream from the shock. The Fermi process, which requires repeated crossings of the shock, is then inhibited. As a result, the post-shock particle spectrum is consistent with a  Maxwellian distribution, for magnetizations $\sigma\gtrsim3\times\ex{3}$ in electron-positron flows and $\sigma\gtrsim\ex{4}$ in electron-ion flows. We have tested that the lack of particle acceleration in high-$\sigma$ shocks is a solid result, confirmed by means of 3D simulations and validated with different choices of the mass ratio (up to $\mime=1600$, close to the realistic value $\mime\simeq1836$).

Weakly magnetized flows ($\sigma\lesssim\ex{3}$ in electron-positron flows; $\sigma\lesssim3\times\ex{5}$ in electron-ion plasmas) are mediated by the Weibel instability, as in unmagnetized shocks \citep{spitkovsky_08,spitkovsky_08b,martins_09,haugbolle_10}. The instability, seeded by the counter-streaming between the incoming flow and a beam of shock-reflected particles that propagate back into the upstream, produces strong small-scale magnetic turbulence that governs the Fermi process. As a result,  the post-shock spectrum in weakly magnetized shocks shows a prominent non-thermal tail of shock-accelerated particles. Regardless of $\sigma$ (in the regime of weakly magnetized flows) or $\gamma_0$ (provided that $\gamma_0\gtrsim10$), the non-thermal tail contains $\sim 1\%$ of particles and $\sim10\%$ of flow energy, and its power-law slope is $p\sim2.5$. In electron-ion shocks, the two species enter the shock nearly in equipartition, as a result of strong energy exchange ahead of the shock mediated by the time-varying Weibel fields. As a result, electrons and ions have similar acceleration efficiencies.

For weakly magnetized shocks, the non-thermal tail of shock-accelerated particles stretches in time to higher and higher energies, following the scaling $\varepsilon_{max}/\gamma_0 m_i c^2\simeq0.5(\ompt)^{1/2}$ in electron-positron plasmas (here, $\omega_{\rm pi}$ indicates the plasma frequency of positrons, equal to the electron plasma frequency $\omega_{\rm pe}$) and $\varepsilon_{max}/\gamma_0 m_i c^2\simeq0.25(\ompt)^{1/2}$ in electron-ion flows (here, $\omega_{\rm pi}=\sqrt{m_e/m_i}\;\omega_{\rm pe}$ is the ion plasma frequency). The relation $\varepsilon_{max}\propto t^{1/2}$, rather than the commonly assumed Bohm scaling $\varepsilon_{max}\propto t$, is a natural consequence of the small-scale nature of the Weibel turbulence \citep{kirk_reville_10,plotnikov_12}. The coefficient of proportionality (i.e., $\simeq0.5$ for electron-positron, $\simeq0.25$ for electron-ion) provides an estimate of the combination $\epsilon_B \lambda_{\comp}$, as shown in \eq{gmax1}. Here, $\epsilon_B$ is the fraction of flow energy converted into magnetic fields in the shock region, and $\lambda_{\comp}$ is the transverse scale of the Weibel modes in units of the plasma frequency $\comp$. We find $\lambda_{\comp}\simeq10$, which implies $\epsilon_B\simeq\ex{2}$ in electron-positron flows and $\epsilon_B\simeq3\times\ex{3}$ in electron-ion plasmas. Our estimates for $\epsilon_B$ do not depend on $\sigma$ (in the regime of weakly magnetized flows) or $\gamma_0$ (provided that $\gamma_0\gtrsim10$).

By measuring the time $t(\varepsilon)$ needed to accelerate the particles up to a given energy $\varepsilon$, our simulations are implicitly providing an estimate for the diffusion coefficient $D\sim c^2 t$ in relativistic perpendicular shocks. We find that $D\simeq 4\, c\,\comp (\varepsilon/\gamma_0 m_i c^2)^2$ in electron-positron flows, and $D\simeq 16\, c\,\comp (\varepsilon/\gamma_0 m_i c^2)^2$ in electron-ion plasmas, regardless of the flow magnetization $\sigma$ or the bulk Lorentz factor $\gamma_0$. This clearly differs from the so-called (and widely used) Bohm scaling, where $D_B\sim c\, r_{L,\sigma}(\varepsilon)\sim \sigma^{-1/2}\,c\,\comp (\varepsilon/\gamma_0 m_i c^2)$. The two estimates differ by a factor of $D/D_B\simeq4\,\sigma^{1/2}(\varepsilon/\gamma_0 m_i c^2)$ in electron-positron flows and $D/D_B\simeq16\,\sigma^{1/2}(\varepsilon/\gamma_0 m_i c^2)$ in electron-ion plasmas. For small $\sigma$, this shows that the Fermi process in relativistic shocks may be even faster than Bohm, if we calibrate the Bohm scaling with the pre-shock field.\footnote{When the Bohm scaling is computed with the self-generated fields, in the formulae above we should replace $\sigma$ with $2 \epsilon_B$.} 

The scaling $\varepsilon_{max}\propto t^{1/2}$ breaks down at an energy $\varepsilon_{sat}$ that depends on the flow magnetization $\sigma$. The thickness of the upstream region populated with Weibel filaments scales as the Larmor radius of the particles at the low-energy end of the non-thermal tail, giving $L_{B,sat}\sim \sigma^{-1/2}\comp$. When the diffusion length of the highest energy particles ($\sim D/c$) reaches $L_{B,sat}$, further evolution to higher energies is inhibited, due to the lack of sufficient magnetic turbulence to oppose advection downstream with the ordered field. At this point, the shock reaches a steady state, and the particle spectrum saturates with an upper energy cutoff $\varepsilon_{sat}/\gamma_0 m_i c^2\simeq 4\,\sigma^{-1/4}$ in electron-positron shocks and $\varepsilon_{sat}/\gamma_0 m_i c^2\simeq 2\,\sigma^{-1/4}$ in electron-ion flows. The saturation in the particle spectrum has been confirmed with 3D simulations. Acceleration to higher energies (beyond $\varepsilon_{sat}$) might be governed by pre-existing upstream turbulence, as argued by \citet{sironi_goodman_07}, provided that sufficient turbulent power exists on scales $\sim L_{B,sat}\simeq10^{13}\sigma_{-9}^{-1/2}\unit{cm}$ (as measured in the post-shock frame).

Our results can provide physically-grounded inputs for models of non-thermal emission from a variety of astrophysical non-thermal sources, with particular relevance to GRB afterglows and the termination shock of PWNe. As we have described in \S\ref{sec:grb}, we find that the acceleration of protons by the external shocks of GRB afterglows is too slow (given that $t\propto \varepsilon^2$) to explain the extreme Lorentz factors of UHECRs. In the early phases of GRB afterglows, electrons can be accelerated up to Lorentz factors $\gamma_{sync,e}\sim10^7$ before suffering catastrophic synchrotron losses in the Weibel-generated fields (as opposed to the background pre-shock field, as argued by \citet{kumar_12}). Their synchrotron radiation can produce the $\sim \unit{GeV}$ photons detected by the Fermi telescope \citep[e.g.,][]{ackermann_10,depasquale_10,ghisellini_10} in early GRB afterglows. A more comprehensive comparison with the observations, including the effect of IC losses \citep[as done by][but without a self-consistent model for the Fermi process in GRB afterglows]{li_waxman_06,piran_nakar_10,barniol-duran_11,li_zhao_11,lemoine_12,sagi_nakar_12}, is deferred to a future study.

As regards to PWNe, Fermi acceleration in the magnetized transverse relativistic shock that terminates the pulsar wind has been believed, for more than 40 years, to be the mechanism behind the conversion of the wind energy into the non-thermal particle spectra inferred in PWNe. In \S \ref{sec:pwn}, we have shown that Fermi acceleration in the  turbulence self-generated at the  termination shock is not able to accelerate the particles to sufficiently high energies to explain the X-ray and gamma-ray emission from the Crab Nebula. To explain the observed spectral signatures of PWNe, we suggest that the turbulence required for the Fermi process might be generated either in the nebula, by large-scale motions \citep{camus_09} cascading down to smaller wavelengths \citep{bucciantini_11}, or at the shock itself, by cyclotron waves \citep{amato_arons_06} excited by a relativistic pair beam that was pre-accelerated in the pulsar wind via magnetic reconnection \citep{arons_12}.


\acknowledgements
L.S. gratefully thanks D.~Giannios and L.~Nava for comments that helped to improve the manuscript. We gratefully thank G.~Pelletier and I.~Plotnikov for fruitful discussions, supported by ISSI.
L.S. is supported by NASA through Einstein
Postdoctoral Fellowship grant number PF1-120090 awarded by the Chandra
X-ray Center, which is operated by the Smithsonian Astrophysical
Observatory for NASA under contract NAS8-03060. A.S. is supported by NSF grant AST-0807381 and NASA grant NNX12AD01G.
The simulations were
performed on the
PICSciE-OIT High Performance Computing Center and Visualization
Laboratory at Princeton University, on XSEDE resources under
contracts No. TG-AST120010 and TG-AST100035, and on NASA High-End Computing (HEC) resources through the NASA Advanced Supercomputing (NAS) Division at Ames Research Center. 


\appendix

\section{Dependence on the Mass Ratio}\label{sec:specmime}
\begin{figure}
\centering
\subfigure[]{
\includegraphics[width=0.48\textwidth]{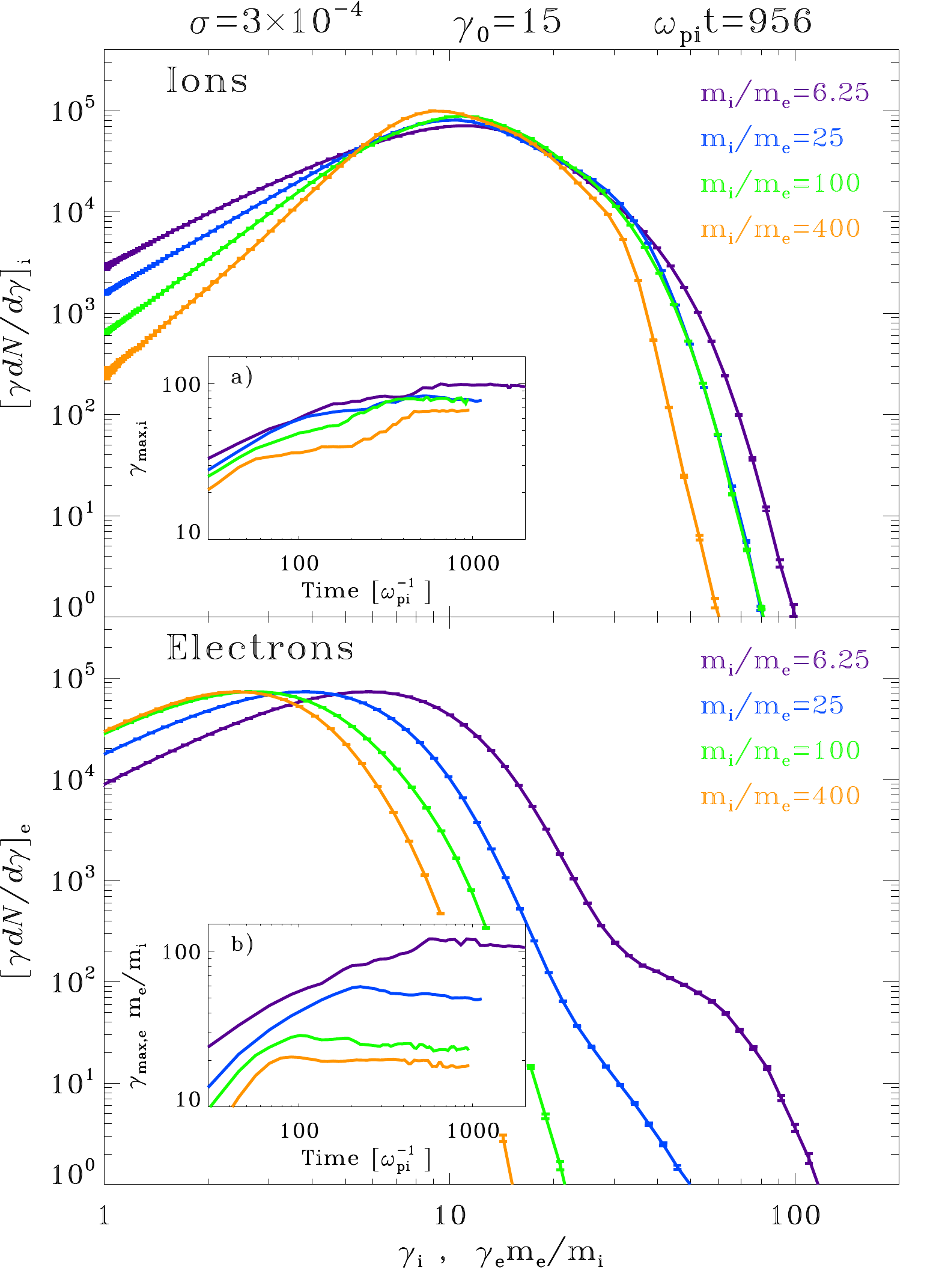}}
\hspace{0.1in}
\subfigure[]{
\includegraphics[width=0.48\textwidth]{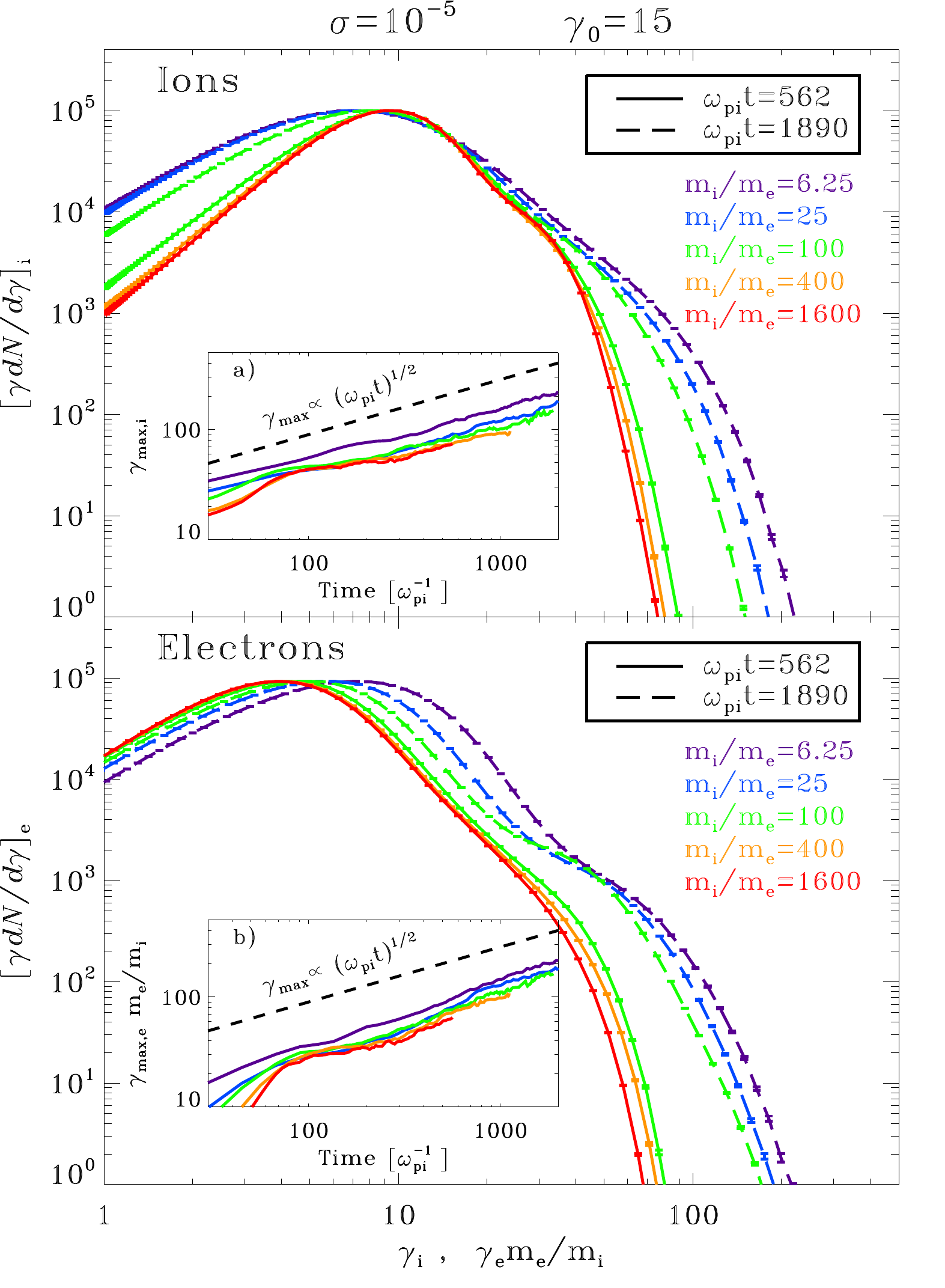}}
\caption{Comparison of downstream particle spectra (upper panel for ions, lower panel for electrons) between different mass ratios, from $\mime=6.25$ up to $\mime=1600$. The left column refers to $\sigma=3\times\ex{4}$, and the right column to $\sigma=\ex{5}$. In the insets, we plot the temporal evolution of the maximum particle energy.}
\label{fig:specmime}
\end{figure}

In \fig{specmime} we analyze the dependence of the post-shock spectrum in electron-ion shocks on the mass ratio $\mime$, up to $\mime=1600$, that approaches the realistic value $\mime\simeq1836$. The left column shows the results for $\sigma=3\times\ex{4}$, and the right column refers to $\sigma=\ex{5}$. We find that the shape of the post-shock spectrum and the temporal scaling of the maximum energy of shock-accelerated particles are nearly insensitive to the mass ratio $\mime$.

For $\sigma=3\times\ex{4}$ (left column in \fig{specmime}), the spectrum of ions (top panel) and of electrons (bottom panel) shows no evidence for particle acceleration, provided that $\mime\gtrsim100$. As discussed in \S\ref{sec:specmi}, the electrons populate a Maxwellian distribution, whereas the ion spectrum is not entirely consistent with a Maxwellian, due to incomplete ion thermalization at the shock front. As the mass ratio decreases (blue line for $\mime=25$, purple line for $\mime=6.25$), the peak of the electron Maxwellian systematically shifts to higher energies, suggesting an increase in the efficiency of ion-to-electron energy transfer ahead of the shock, for smaller $\mime$. Yet, the location of the thermal peak in the electron spectrum only varies by a factor of $\sim3$, as the mass ratio changes from $\mime=6.25$ up to $\mime=400$. Efficient electron heating ahead of the shock requires strong Weibel turbulence, which can be excited if the upstream electrons are unmagnetized. This requires that the initial electron magnetization $\sigma_{e,0}=(m_i/m_e)\,\sigma$ should not be too large, a constraint that, for fixed $\sigma$, can be  satisfied more easily for smaller $\mime$. This explains the trend in the electron thermal peak seen in the bottom panel of \fig{specmime} (left column). 

The increase in the number of non-thermal electrons as the mass ratio decreases (bottom left panel of \fig{specmime}) is then a consequence of the heating experienced by the incoming electrons in low-$\mime$ shocks. If the electrons are entering the shock with a larger energy, they are more likely to be injected into the acceleration process, which explains the trend in the normalization of the electron non-thermal tails seen in the bottom left panel of \fig{specmime}. Yet, even for  $\mime=6.25$, namely the case that shows the largest acceleration efficiency, the maximum electron Lorentz factor stops growing after $\ompt\sim600$ (purple line in the inset (b) of the bottom left panel). 

For $\sigma=\ex{5}$ (right column in \fig{specmime}), the post-shock spectrum of both ions (top panel) and electrons (bottom panel) shows a prominent non-thermal tail, extending in time to higher and higher energies. In the right column of \fig{specmime}, we plot the spectra at $\ompt=562$ with solid lines, and at $\ompt=1890$ with dashed lines. For both times, the ion and electron spectra are nearly identical across a wide range of mass ratios: at relatively early times ($\ompt=562$), we can follow the physics of electron-ion shocks with large mass ratios ($\mime=100$ in green, $\mime=400$ in yellow, and the nearly realistic case $\mime=1600$ in red); when evolving the shock to longer times  ($\ompt=1890$), we are limited to relatively small mass ratios ($\mime=6.25$ in purple, $\mime=25$ in blue, $\mime=100$ in green). The shape of the spectra (for both ions and electrons) is remarkably similar across the whole range of mass ratios we explore, with the only difference being a moderately larger degree of electron heating for smaller $\mime$, as explained above for $\sigma=3\times\ex{4}$ shocks. Yet, the difference in the electron peak energy between $\mime=6.25$ and $\mime=100$ at $\ompt=1890$ is less than a factor of two, and it is even smaller between $\mime=100$ and $\mime=1600$ at $\ompt=562$. 

As time progresses, the non-thermal tail of both ions (top panel in the right column of \fig{specmime}) and electrons (bottom panel) systematically extends to higher energies, as exemplified by the case with $\mime=100$ (compare the solid green lines for $\ompt=562$ to the dashed green lines for $\ompt=1890$). The temporal evolution of the maximum ion and electron Lorentz factor is plotted in the insets of the right column of \fig{specmime} (inset (a) for ions and (b) for electrons), showing that the scaling as $\propto(\ompt)^{1/2}$ is realized at late times independently of the mass ratio. Based on this evidence, we argue that the results presented in the main body of the paper, both in terms of acceleration efficiency and of energization rate, are going to hold even for the realistic mass ratio $\mime\simeq1836$. 

\section{Dependence on the Field Orientation Relative to the Simulation Plane}\label{sec:specphi}
In the main body of the paper, we have preferentially employed 2D computational boxes with the pre-shock magnetic field $B_0$ oriented perpendicular to the simulation plane (i.e., out-of-plane fields). Here, we demonstrate that this configuration is in excellent agreement with 3D simulations.

\begin{figure}
\centering
\subfigure[]{
\includegraphics[width=0.49\textwidth]{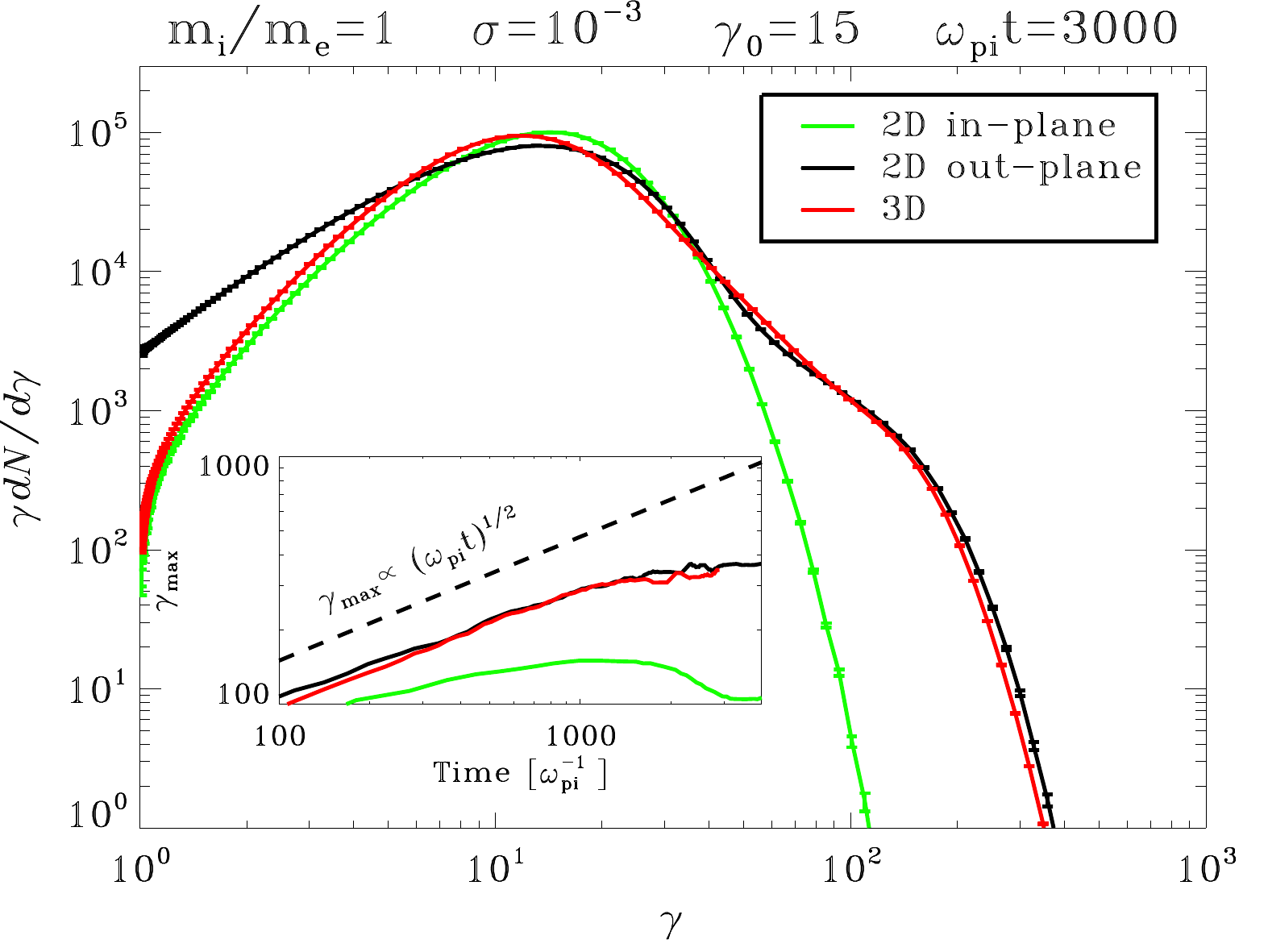}}
\hspace{0.1in}
\subfigure[]{
\includegraphics[width=0.47\textwidth]{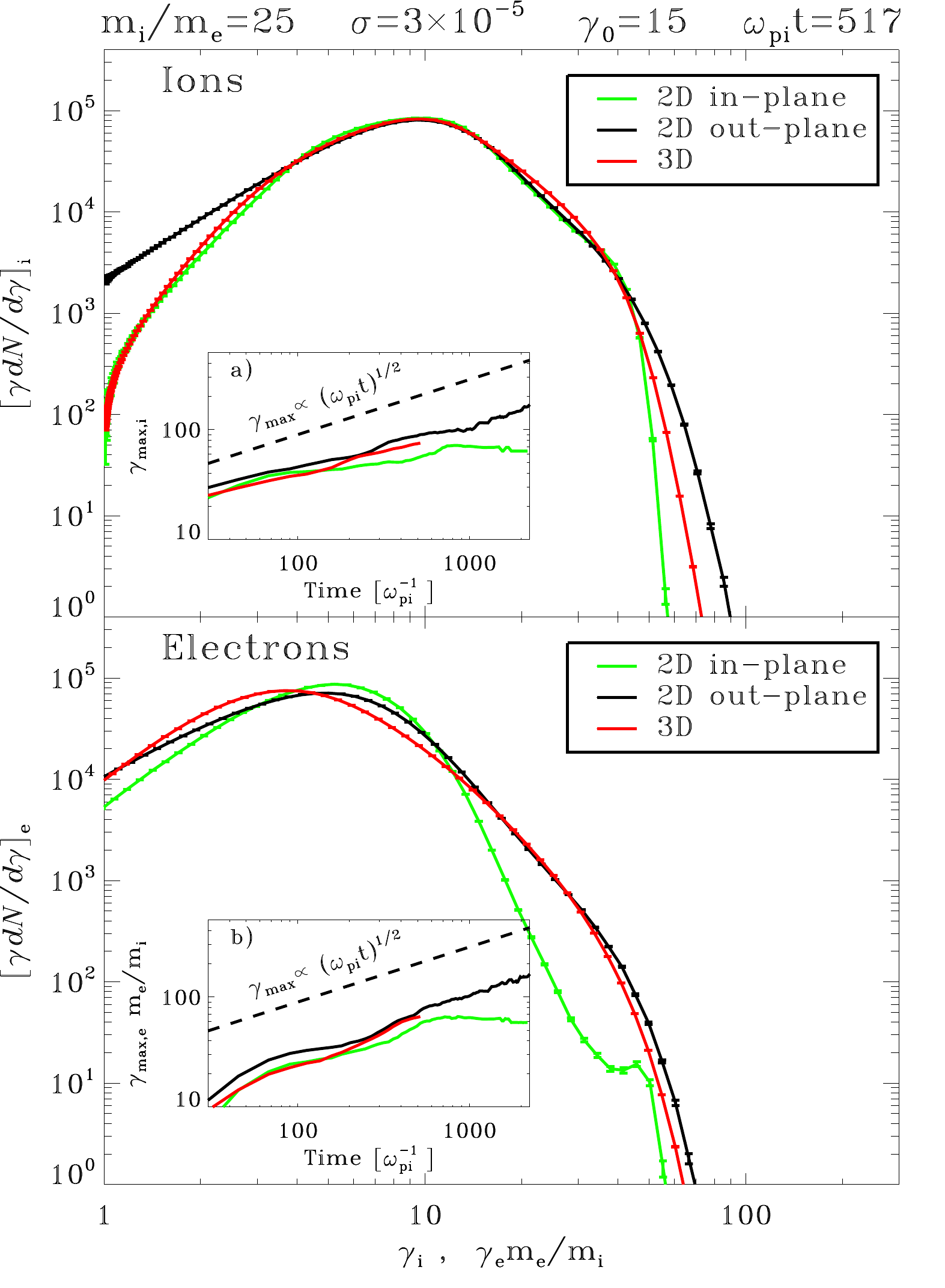}}
\caption{Comparison of downstream particle spectra among different geometries of the upstream background magnetic field. We compare  2D simulations with in-plane fields (green), 2D simulations with out-of-plane fields (black; this is the configuration adopted in the main body of the paper), and 3D simulations (red). The left column refers to an electron-positron shock with magnetization $\sigma=\ex{3}$, whereas the right column shows the ion and electron spectra of an electron-ion shock ($\mime=25$) with $\sigma=3\times\ex{5}$. In the insets, we plot the temporal evolution of the maximum particle energy.}
\label{fig:specphi}
\end{figure}

In \fig{specphi}, we show the post-shock spectrum for different geometries of the upstream background magnetic field. We compare  2D simulations with in-plane fields (green), 2D simulations with out-of-plane fields (black), and 3D simulations (red), for both electron-positron flows with $\sigma=\ex{3}$ (left column) and electron-ion flows with $\mime=25$ and $\sigma=3\times\ex{5}$ (right column). The disagreement between 3D simulations and 2D runs with out-of-plane fields at low energies (below the thermal peak) is a mere consequence of the different adiabatic index, which changes the analytic formula of a Maxwellian distribution. In fact, for 2D simulations with out-of-plane fields, the plasma is constrained to move in the plane orthogonal to the field, which coincides with the simulation plane. In contrast, in 2D runs with in-plane fields, these two planes are not degenerate, the particle velocities can sample all three dimensions, and the adiabatic index is the same as in 3D (which explains the agreement at low energies between the green and red lines).

At high energies (i.e., in the non-thermal tail), 2D simulations with out-of-plane fields are in good agreement with 3D simulations, whereas 2D runs with in-plane fields systematically underestimate the acceleration efficiency.  3D simulations and 2D runs with out-of-plane fields yield consistent results as regards to both the acceleration efficiency and the temporal evolution of the maximum energy of accelerated particles (see the insets in \fig{specphi}). In summary, the 2D simulations with out-of-plane fields employed in the main body of the paper provide an excellent description of the 3D acceleration physics of relativistic perpendicular shocks. 

The artificial suppression of the acceleration efficiency in 2D simulations with in-plane fields may be explained by Jones' theorem \citep{jones_98}, that charged particles cannot move farther than one Larmor radius from the plane defined by the background magnetic field (along $\bmath{\hat{y}}$) and the $\bmath{\hat{z}}$ direction orthogonal to the simulation domain. Equivalently, we could argue that the diffusion of particles back into the upstream, that is required for injection into the Fermi process, is likely to be suppressed in the case of in-plane fields, since the growth of the Weibel magnetic fields (preferentially along  $\bmath{\hat{z}}$, for 2D runs) cannot efficiently compete with advection by the ordered background field (directed along $\bmath{\hat{y}}$). In contrast, if the pre-shock field is initialized orthogonal to the simulation plane, its effects can be  counteracted more easily by the Weibel modes, and injection into the Fermi process would be facilitated.

Finally, we remark that the difference seen in \fig{specphi} between the acceleration capabilities of 2D shocks with in-plane and out-of-plane fields is most dramatic for magnetizations close to the transition between poorly-accelerating and efficiently-accelerating shocks (this is what motivated the choice of $\sigma$ in \fig{specphi}). For magnetizations that are a factor of $\sim3$ larger (or smaller) than this critical boundary, the post-shock spectra in 2D simulations are essentially the same regardless of the field orientation relative to the simulation plane.

\bibliography{max}
\end{document}